\pdfoutput=1
\documentclass[twocolumn,english,aps,longbibliography,superscriptaddress,prb]{revtex4-1}

\usepackage{babel}
\usepackage{amsmath}
\usepackage{amssymb}
\usepackage{graphicx}
\usepackage{xcolor}
\usepackage{tikz}

\definecolor{bleu}{rgb}{0.00,0.45,0.74}
\definecolor{jaune}{rgb}{0.93,0.69,0.13}
\definecolor{rouge}{rgb}{0.85,0.33,0.10}
\definecolor{vert}{rgb}{0.47,0.67,0.19}
\definecolor{violet}{rgb}{0.49,0.18,0.56}

\begin{document}

\title{Tensor network investigation of the hard-square model}

\author{Samuel Nyckees}
\affiliation{Institute of Physics, Ecole Polytechnique F\'ed\'erale de Lausanne (EPFL), CH-1015 Lausanne, Switzerland}
\author{Fr\'ed\'eric Mila}
\affiliation{Institute of Physics, Ecole Polytechnique F\'ed\'erale de Lausanne (EPFL), CH-1015 Lausanne, Switzerland}

\date{\today}
\begin{abstract} 
Using the corner-transfer matrix renormalization group to contract the tensor network that describes its partition function, we investigate the nature of the phase transitions of the hard-square model, one of the exactly solved models of statistical physics for which Baxter has found an integrable manifold. The motivation is twofold: assess the power of tensor networks for such models, and probe the 2D classical analog of a 1D quantum model of hard-core bosons that has recently attracted significant attention in the context of experiments on chains of Rydberg atoms. Accordingly, we concentrate on two planes in the 3D parameter space spanned by the activity and the coupling constants in the two diagonal directions. We first investigate the only case studied so far with Monte Carlo simulations, the case of opposite coupling constants. We confirm that, away and not too far from the integrable 3-state Potts point, the transition out of the period-3 phase appears to be unique in the Huse-Fisher chiral universality class, albeit with significantly different exponents as compared to Monte Carlo. We also identify two additional phase transitions not reported so far for that model, a Lifshitz disorder line, and an Ising transition for large enough activity. To make contact with 1D quantum models of Rydberg atoms, we then turn to a plane where the ferromagnetic coupling is kept fixed, and we show that the resulting phase diagram is very similar, the only difference being that the Ising transition becomes first-order through a tricritical Ising point, in agreement with Baxter's prediction that this plane should contain a tricritical Ising point, and in remarkable, almost quantitative agreement with the phase diagram of the 1D quantum version of the model. 

\end{abstract}

\maketitle

%%%%%%%%%%%%%%%%%%%%%%%%%%%%%%%%%%%%% INTRODUCTION %%%%%%%%%%%%%%%%%%%%%%%%%%%%%%%%%%%%

\section{Introduction}

Commensurate-incommensurate (C-IC) transitions have recently attracted renewed attention due to their experimental realisation in Rydberg atoms\cite{lukin2017,lukin2019}. Their nature has long been debated and studied in both classical\cite{cardy,AUYANG1996,yeomans1985,Selke1982,Duxbury,sato,houlrik1986,auyang1987,baxter1988,schulz1980,SelkeExperiment} and quantum systems\cite{howes1983,fendley,chepiga_mila_PRL,samajdar,Everts_1989,hughes,HOWES1983169,sachdev_dual,CENTEN1982585}. The problem was initially introduced in the context of adsorbed monolayers\cite{Ostlund,Huse1981,schulz,HuseFisher1984}. The physics at C-IC transitions is controlled by domain walls and dislocations, and it becomes a subtle problem when  walls between domains $A\mid B$ and $B\mid A$ have different energies. Their average distance defines the pitch or wave-vector $q$, which goes to the commensurate value with a power law described by the critical exponent $\bar{\beta}$ ($q-q_0\sim t^{\bar{\beta}}$). Based on scaling arguments, Huse and Fisher\cite{HuseFisher1982} first proposed the existence of a unique transition for $p=3,4$ which would be characterised by the fact that the product of the wave-vector along the incommensurate direction with the correlation length goes to a strictly positive constant at criticality, $\xi (q -q_0)\rightarrow{} cst>0$ (or equivalently by the fact that $\nu_\parallel = \bar{\beta}$, where $\nu_\parallel$ characterises the power law divergence of the correlation length along the incommensurate direction). This contrasts with the usual isotropic transitions for which such a product is believed to go to zero at the critical temperature ($\bar{\beta}>\nu_\parallel$). Studies treating the dislocations perturbatively have shown that for $p>2$, the transition can also take place through a two-step process separated by a floating phase: first through a Pokrovsky-Talapov (PT)\cite{Pokrovsky_Talapov} transition characterised by critical exponents $(\nu_x, \nu_y, \bar{\beta} )=(1/2,1,1/2)$ at low temperature, then through a Kosterlitz-Thouless (KT)\cite{Kosterlitz_Thouless_1973} transition characterised by the exponential divergence of the correlation length coming from high temperature\cite{Den_Nijs}. For a two-step transition, the product $\xi (q-q_0)$ thus diverges approaching the floating phase, and this leaves three different scenarios for the C-IC transition which can all be distinguished by the behavior of the product $\xi (q-q_0)$.

Originally introduced by Baxter\cite{baxter1980,Baxter1981}, the hard-square model is one of the paradigmatic models to study this issue because it hosts commensurate melting from period-2 and period-3 phases, and because it contains an integrable manifold inside which transitions have been fully characterized by Baxter: the melting of the $2\times1$ phase occurs via an Ising tricritical transition while the $3\times1$ phase melts through a 3-state Potts transition. Away from the 3-state Potts points, Huse\cite{Huse_1983} argued that a chiral perturbation is present, and that the transition has to change nature and could become chiral. Since away from the 3-state Potts point the model is not integrable along the transition line, the only way to test Huse's prediction is to resort to numerical approaches. This has been attempted with Monte Carlo
simulations by Bartelt \textit{et al} in the late eighties, for the model with diagonal and anti-diagonal interactions respectively attractive and repulsive and of the same intensity. The results are consistent with a chiral transition close to the Potts point, with an exponent $\bar{\beta}\simeq 0.8$. This exponent disagrees with later results on other models\cite{cardy,Nyckees2020} and with experimental results on reconstructed surfaces\cite{abernathy}, which all point to an exponent $\bar{\beta}=2/3$. 

In the present paper, we revisit the hard-square model using the corner transfer matrix renormalization group (CTMRG), a method introduced in the mid-nineties by Nishino and Okunishi\cite{nishino} and used recently on the chiral Potts\cite{Nyckees2020} and Ashkin-Teller\cite{Nyckees2022} models. As we shall see, this approach confirms and complements the Monte Carlo investigation by Bartelt \textit{et al} \cite{bartelt} of the model with opposite diagonal and anti-diagonal interactions, with in particular an estimate of $\bar{\beta}\simeq 2/3$ in better agreement with other results, and the identification of a disorder line and an Ising transition at larger activity and temperature. We also study another cut through the parameter space of the hard-square model that corresponds to the 2D classical version of a 1D bosonic quantum model recently studied in the context of chains of Rydberg atoms\cite{fendley,chepiga_mila_PRL,samajdar}, with a phase diagram in excellent agreement with its 1D counterpart.

The paper is organised as follows. In Section II we describe the model and recall some of the exact results and previous work. In Section III, we present our main results for the  model along the cut initially studied with Monte Carlo, as well as the phase diagram for the other cut that corresponds to the 1D quantum bosonic model. The results are put in perspective in Section IV. The technical aspects of the method, which has already been used for other models\cite{Nyckees2020,Nyckees2022}, are recalled in the appendices, as well as the mapping between the 1D quantum bosonic model and the hard-square model.

\section{The model}
\begin{figure}[t!]
\centering
\includegraphics[width = 0.45\textwidth]{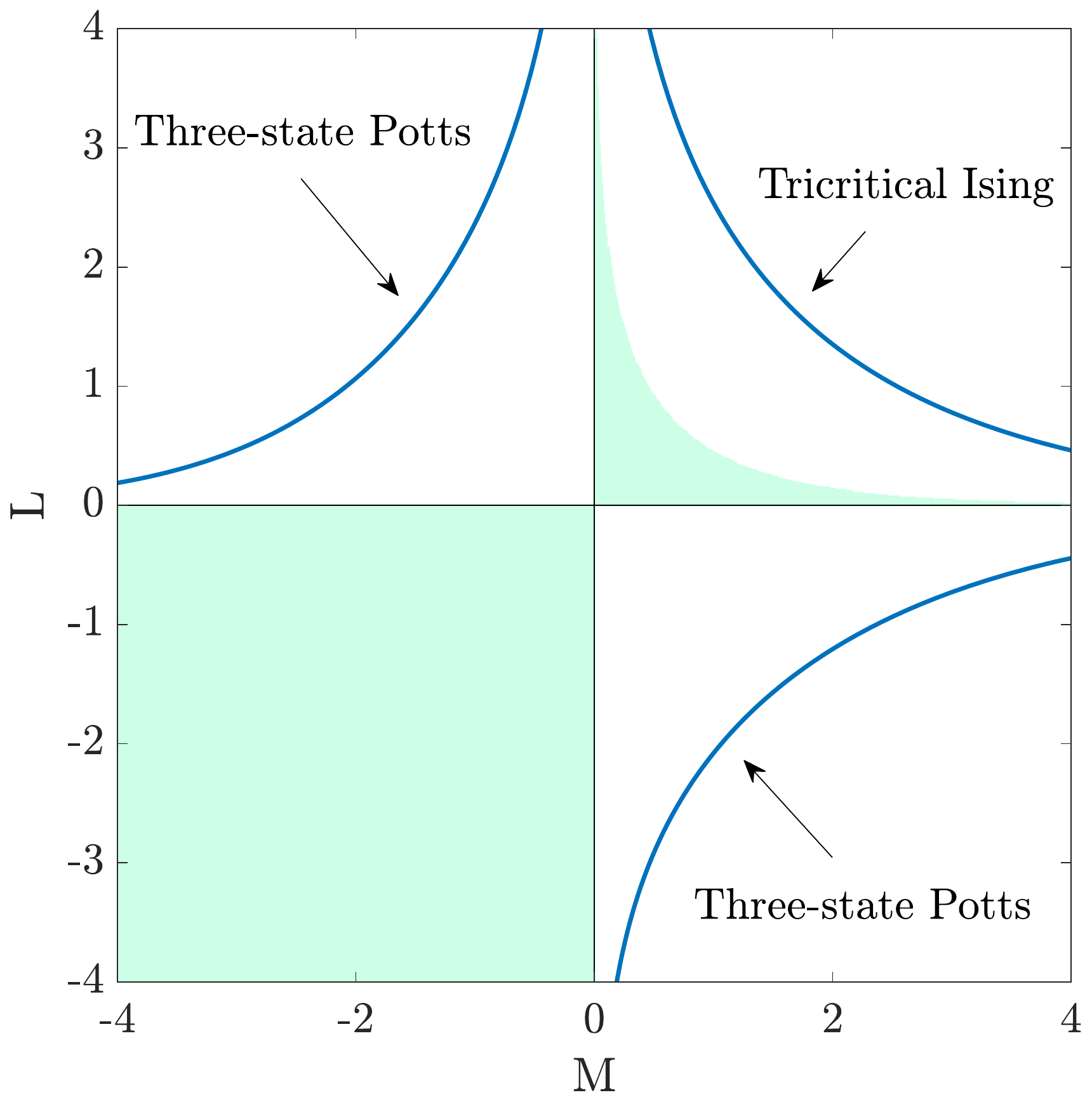}
\caption{Phase diagram of the model on the integrable manifold as derived by Baxter. The $z$ value on this manifold is not constant but is given by Eq.\ref{eqn:Int1}. The shaded area corresponds to the non-physical parameter range $z<0$. The cut with opposite interactions on the diagonals and the antidiagonals corresponds to $M = -L>0$ and has only one integrable phase transition in the 3-state Potts universality class, while a cut $L = cst>0$ has an additional integrable Ising tricritical point.}
\label{fig:Exact}
\end{figure}

The hard-square model with diagonal interactions is defined on a square lattice with spins on the vertex taking value $n\in\{1,0\}$ . If $n=1$, the spin is said to be filled while if $n=0$ the spin is said to be empty. The model is defined in the grand canonical ensemble by
\begin{align}
    \beta H = - M\sum_{x,y} n_{x,y}n_{x+1,y+1} -L \sum_{x,y} n_{x,y} n_{x+1,y-1} 
\end{align}
with $\beta$ the inverse temperature and $M = \beta J_1$ and $L= \beta J_{2}$ where $J_1$ and $J_{2}$ are the respective diagonal and anti-diagonal coupling constants. The hard-core constraint forbids two neighbouring spins from both be filled, leading to the partition function:
\begin{align}
    Z =\sum_{\{n\}} \prod_{\langle i, j\rangle} (1-n_in_j) e^{-\beta H} \prod_i z^{n_{i}}
\end{align}
where the activity is defined as $z = e^{\mu}$, with $\mu$ usually referred to as the chemical potential. Baxter showed that there exists an integrable surface in the three dimensional manifold $(z,M,L)$ parametrised by
\begin{align}
    z = (1-e^{-L})(1-e^{-M})/(e^{L+M} - e^{L} -e^{M}).
    \label{eqn:Int1}
\end{align}
On this manifold, the phase transitions occur at:
\begin{align}
   \frac{z }{(1-z e^{L+M})^2}= \frac{1}{2}(11+5\sqrt{5}).
   \label{eqn:Int2}
\end{align}
The solutions for which $M,L>0$ were shown\cite{HuseHardSquare} to be Ising tricritical points, while solutions for which $M>0, L<0$ or $M<0, L >0$ belong to the three-state Potts universality class described by the critical exponents $(\nu, \alpha, \bar{\beta})=(5/6,1/3,5/3)$. The projection of these lines onto the ($M$,$L$) plane are shown in Fig.\ref{fig:Exact}.

The critical density is also known and given by:
\begin{align}
    \rho_c & = (5-\sqrt{5})/10 \simeq 0.27639.
\end{align}
More recently, Sachdev and Fendley\cite{fendley} revisited the model through its one dimensional quantum equivalent Hamiltonian defined by
\begin{align*}
 H = \sum_{i} -\omega (\hat{d}_j + \hat{d}_j^\dagger) + U \hat{n}_j + V \hat{n}_{j-1}\hat{n}_{j+1}
\end{align*}
with the constraints $\hat{n}_i \hat{n}_{i+1} = 0$ and $\hat{n}_i(\hat{n}_i -1)=0$. The 1+1 correspondence is done via the transfer matrix formalism. One recovers the classical partition function from the quantum Hamiltonian in the infinite anisotropic limit. More precisely, one needs to take the diagonal transfer matrix in the $L \rightarrow \infty$ and $z, M\rightarrow 0$ limits in such a way that 
\begin{align}
\frac{V}{\omega} & = - M e^{L/2}, \qquad \frac{U}{\omega} = e^{L/2}\zeta
\label{eq:EqSF}
\end{align}
are kept constant, with $\zeta = 1-ze^{L}$. The $x+y$ direction then plays the role of the time direction and the hard-core constraint translates into $\hat{n}_j \hat{n}_{j+1} = 0$ while the constraint $\hat{n}_j(1-\hat{n}_j) = 0$ is due to the spin taking value into $\{0,1\}$. We illustrate the mapping in Fig. \ref{fig:corresp}. More details on how such a correspondance is established are given in Appendix D.

We note that throughout the whole study, due to practical reasons explained in the appendices, we are only able to measure the correlation length and the wave vector along the $x$ and $y$ directions. This is unfortunate since the commensurate direction lies along the $x+y$ axis, but one can live with this restriction, as we now explain. Indeed, if the transition is conformal with anisotropic exponent $\nu_x/\nu_y = 1$ as for the three-state Potts point, the direction along which the correlation length and wave vectors are measured does not matter and one always recovers its critical exponents. Now, if the transition is anisotropic along the $x\pm y$ directions with $ \nu_{x-y}\neq \nu_{x+y}$,  the analysis of the correlation length in the $x$ or $y$ direction will both give the same critical exponent $\nu = \min(\nu_{x-y}, \nu_{x+y})$, which we expect to be equal to $\nu_{x-y}$ if the incommensurate correlations are in the $x-y$ direction from our experience with other models. Besides, $\bar{\beta}_{x+y}$ is not defined due to $q$ being strictly constant everywhere along the $x+y$ direction. This in turn gives $\bar{\beta}_{x-y} = \bar{\beta}_x = \bar{\beta}_y$ and the investigation of $\bar{\beta}$ will not be hampered. This means that it will be possible to check the criterion for a chiral transition: $\bar{\beta}_{x-y} = \nu_{x-y}^{LT}$. The only thing that will not be directly accessible is the dynamical exponent $z=\nu_{x+y}/\nu_{x-y}$, but we can get information on it with hyperscaling and an estimate of the specific heat exponent $\alpha$. More details are provided in the appendices.

 \begin{figure}[t!]
\centering
\includegraphics[width = 0.45\textwidth]{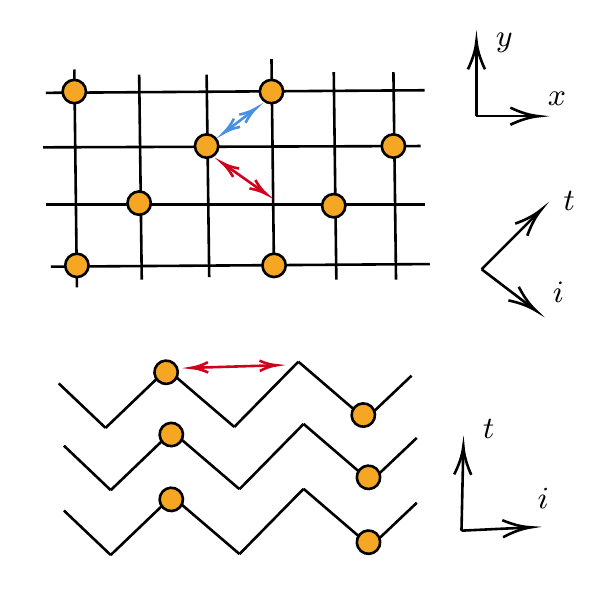}
\caption{Correspondence between the hard-core boson and the hard-square model in the $3\times1$ phase with the red and blue arrows respectively representing the repulsive and attractive interactions. }
\label{fig:corresp}
\end{figure}

\section{Results}

\subsection{Ferro-antiferromagnetic case}

\begin{figure}[t!]
\centering
\includegraphics[width = .45\textwidth]{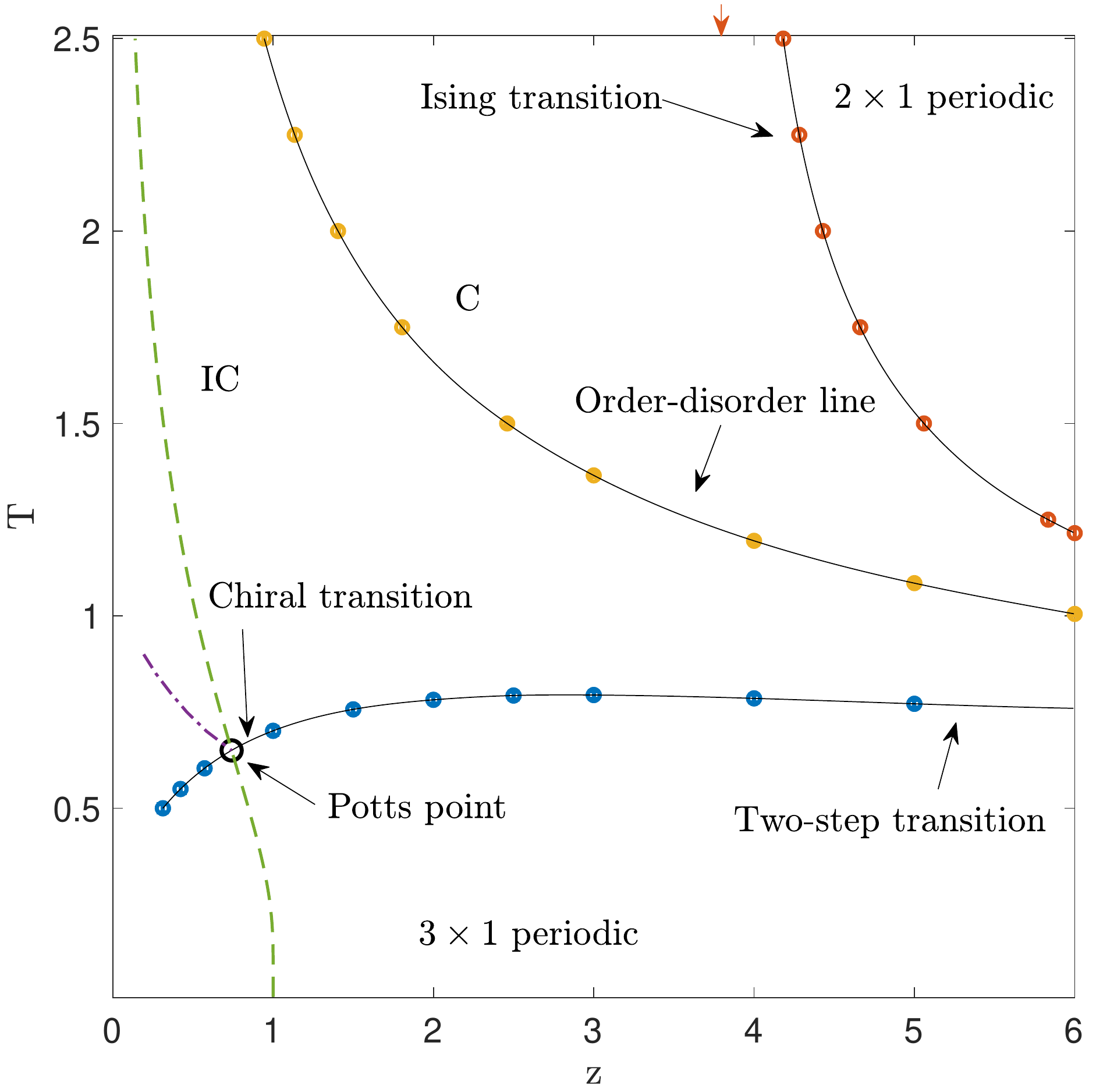} 
\caption{Phase diagram of the model for $M = -L$ where we have plotted:  (\protect\tikz \protect\draw[black] (0,0) circle (.5ex);)  the three-State Potts point, (\protect\tikz \protect\draw[rouge] (0,0) circle (.5ex);)  the Ising transition, (\protect\tikz \protect\draw[bleu] (0,0) circle (.5ex);)  the chiral / PT transitions,(\protect\tikz \protect\draw[jaune] (0,0) circle (.5ex);) the order-disorder line, ( \protect\tikz \protect\draw[thick, vert,-] (0,0.1) to (0.2,0.1) ; ) the  integrable line,  ( \protect\tikz \protect\draw[thick, violet,-] (0,0.1) to (0.2,0.1) ; ) and the $2\pi/3$ commensurate line that terminates at the Potts point. The red arrow at the top represents the hard-core lattice gas critical activity. Black lines are guides to the eyes. }
\label{fig:PhaseDiagram}
\end{figure}

We first address the $M = -L$ case with $J_1 = 1$ and  $J_2 = -1$, which amounts of having attractive and repulsive interactions in the $x+y$ and $x-y$ directions respectively, favouring sites on the diagonal to either be all full or all empty while the sites on the anti-diagonal will have a filling $1/3$. Eq. \ref{eqn:Int2} with $M=-L$ has only one solution for which $M>0$ and $L<0$, thus belonging to the three-state Potts universality class. This cut of the three-dimensional phase diagram has already been studied with Monte Carlo \cite{bartelt} in the vicinity of the incommensurate - $2\pi/3$-commensurate transition. We complete the study by mapping the whole two-dimensional phase diagram and present the results in Fig. \ref{fig:PhaseDiagram}.

We note that in the infinite temperature limit for $M = -L$, we recover the hard-core square lattice gas, which transition occurs at $z_c = 3.7962\pm 0.0001$\cite{Baxter1980LG} and is believed to belong to the Ising universality class\cite{Guo2002}. We expect this transition to persist at finite temperature, and to either stay in the Ising universality class or to become a first order transition at an Ising tricritical point.

\subsubsection{Benchmark : Three-State Potts point}

We now turn to the investigation of the phase diagram and benchmark our algorithm on the three-state Potts points, whose exact location is known and given by $(z_c, T_c) \simeq ( 0.7414,  0.6504)$. We found from the ordered and disordered phase $\nu = 0.832\pm 0.001$ and $\nu = 0.828\pm 0.002$   respectively, in good agreement with the exact result 5/6. We further measure $\bar{\beta}_x = 1.63 \pm 0.01$ and $\bar{\beta}_y = 1.64\pm 0.01$, also in reasonable agreement with the theoretical value 5/3. We note that due to a better extrapolation with respect to the gaps of the transfer matrix, the exponents $\bar{\beta}_y$ obtained at a given temperature have smaller error bars than exponents $\bar{\beta}_x$. Thus, from now on, we will focus on $\bar{\beta}_y$ and $\nu_y$ rather than $\bar{\beta}_x$ and $\nu_x$.

\begin{figure}[t!]
\centering
\includegraphics[width = 0.45\textwidth]{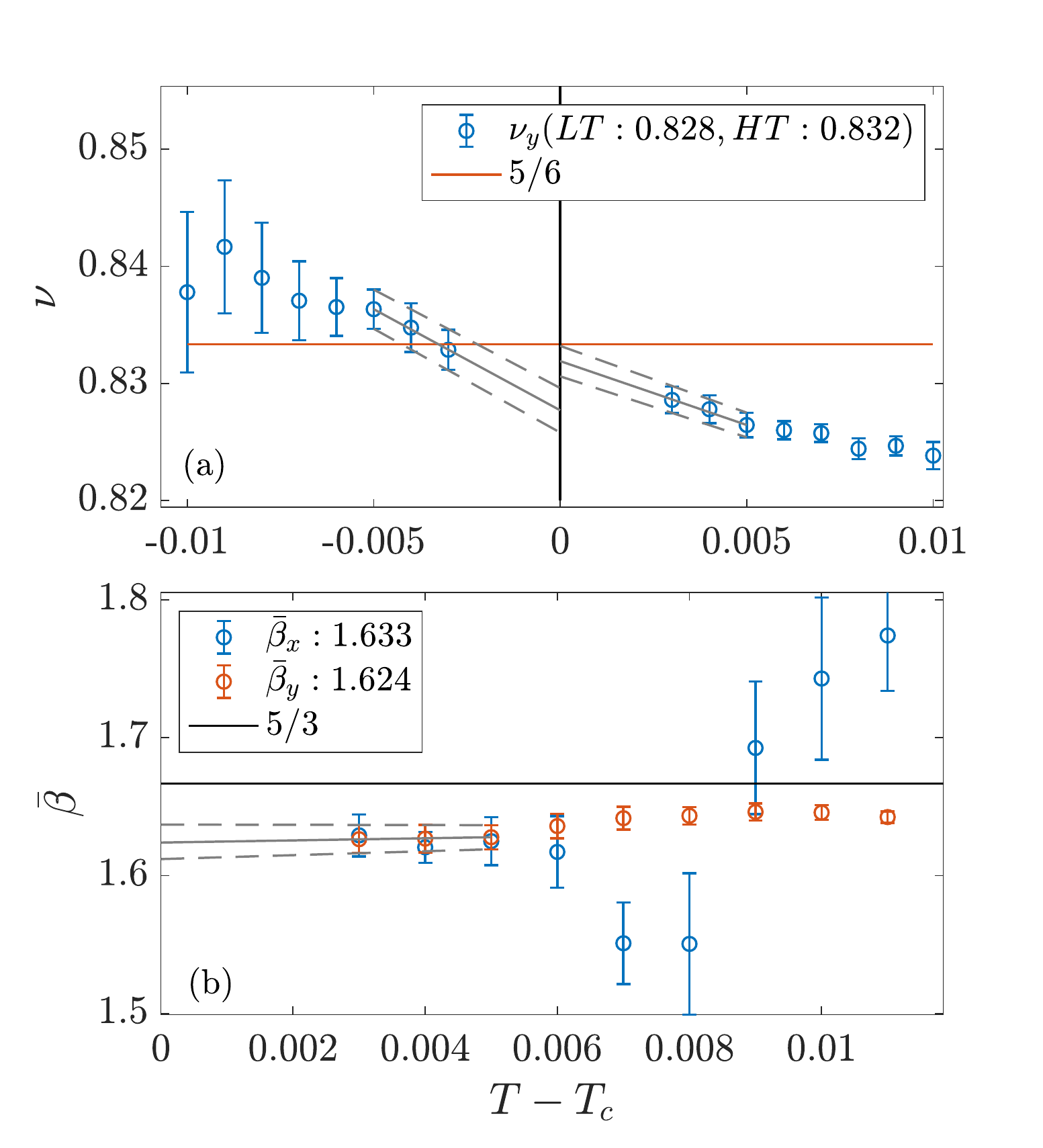}
\caption{Effective exponent $\bar{\beta}$ (b) and $\nu_y$ (a) when approaching the Potts point. We do not show $\nu_x$ since it is exactly equal to $\nu_y$ by symmetry. Error bars and extrapolated exponents are computed with a linear fit (dashed line) on the last three points. For the sake of clarity we do not show the extrapolation on $\bar{\beta}_x$. Simulations were done with $\chi\in[100,200]$. }
\label{fig:Pottsbm}
\end{figure}
% LT nu
% 0.8277 		0.828
% 0.8296		0.830	
% 0.8258		0.826 err 0.002
%HT
% 0.8319		0.832
% 0.8332		0.833
% 0.8306		0.831.  err 0.001
% dqy 
% 1.637,           0.013
% 1.624, 
% 1.612
%dqx 
% 1.633       err 0.011
% 1.644
% 1.623

\subsubsection{Ising transition}

\begin{figure}[t!]
\centering
\includegraphics[width = 0.45\textwidth]{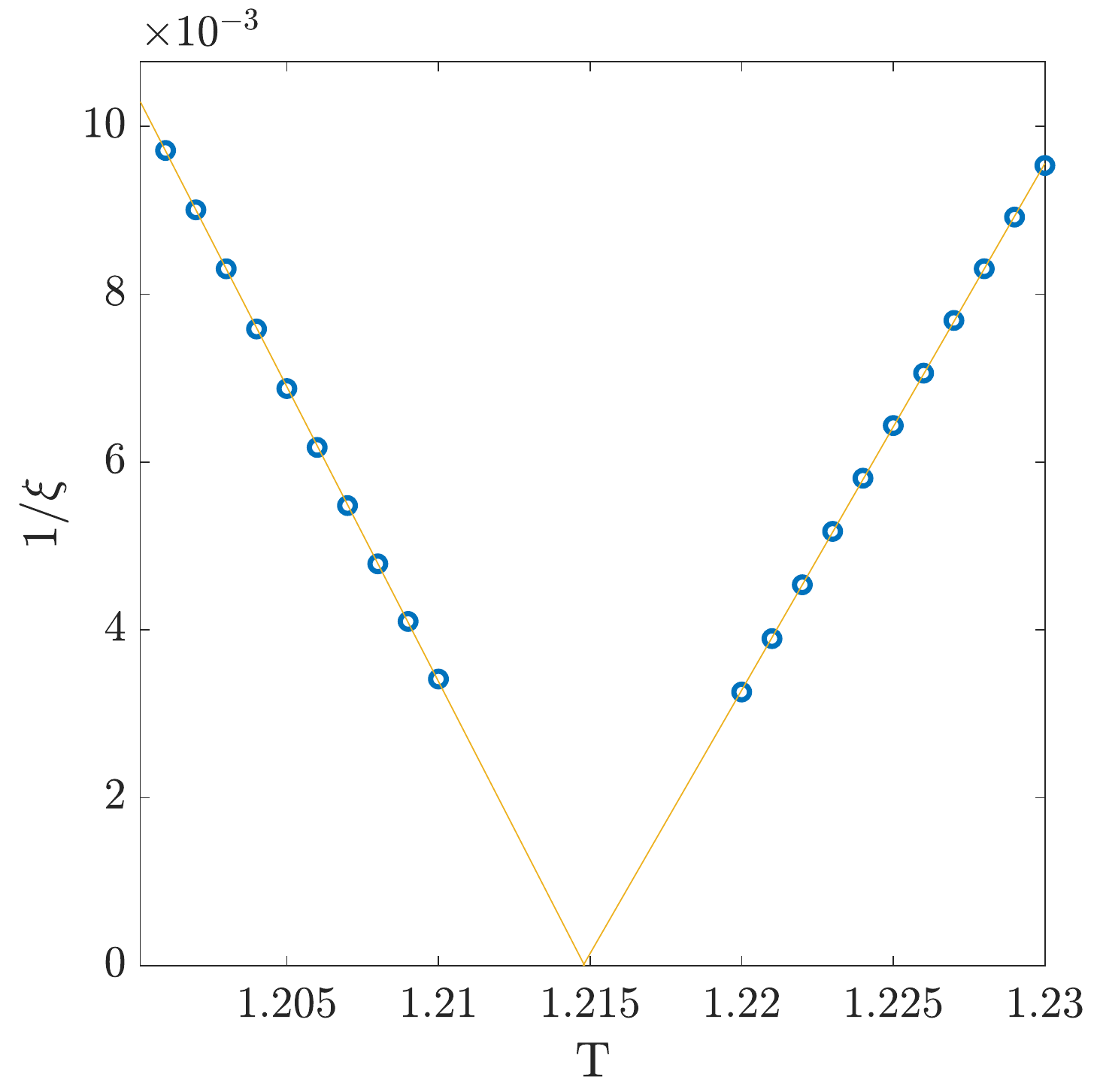}
\caption{Inverse correlation length as a function of temperature at $z=6$. Linear fits from both sides of the transition give two critical temperatures that only differ from each other by about $5 \cdot 10^{-5}$.}
\label{fig:Ising}
\end{figure}

At finite temperature, and in the high chemical potential limit, one enters a $\pi$-commensurate phase in which only two types of domains walls are possible. The transition is thus expected to either be first order or to belong to the Ising universality class. The numerical evidence is clearly in favour of Ising. In particular, at $z = 6$, we found that the inverse correlation length from both sides of the transition goes to zero linearly, in agreement with critical exponent $\nu = 1$. In that case, the critical temperature can be fixed by the intersection from a linear fit of the inverse correlation length with the temperature axis. We can see on Fig. \ref{fig:Ising} that such fits intersect the temperature axis at the same point from both sides of the transition in agreement with a unique critical temperature. We found similar results all along the transition and have studied it up to the largest temperature, $T = 2.5$. If we recall that in the infinite temperature limit one recovers the hard-square lattice gas which transition is believed to belong to the Ising universality class as well, then we can conclude that the transition from the infinite temperature up to at least $z = 6$ belongs to the Ising universality class, and that if in the large activity limit it becomes a first order one, the tricritical point would be located at $z>6$.

% left intersection   T1 = 1.2147710
% right intersection T2 = 1.2148205

\subsubsection{Lifshitz C-IC transition}

\begin{figure}[t!]
\centering
\includegraphics[width = 0.45\textwidth]{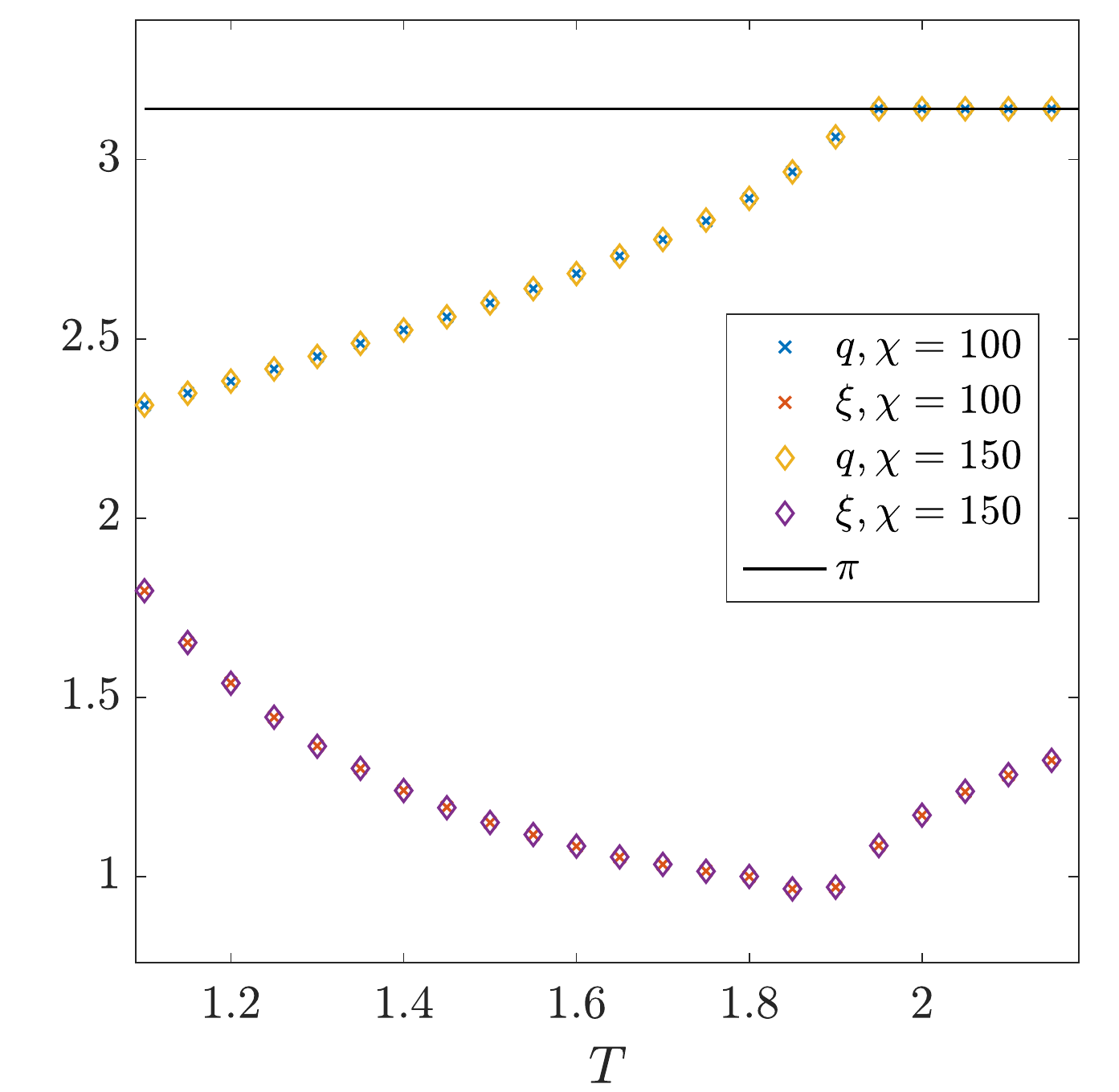}
\caption{Correlation length and wave-vector as function of temperature at $z = 1.52$. The black line represents the $\pi$ constant line. We can observe a $\pi$ commensurate - incommensurate transition around $T = 1.9$.}
\label{fig:LifshitzTrans}
\end{figure}

We found the IC and $\pi$-commensurate phases to be separated by a disorder line of the first kind\cite{schollwoeck_bilbiq}. Such transitions are characterised by an asymmetric temperature dependence of the correlation length, with infinite slope from the commensurate side and a finite derivative from the incommensurate one. We present in Fig. \ref{fig:LifshitzTrans} the results obtained along the $z = 1.52$ cut, where these features can clearly be observed. Simulations were done for finite bond dimension $\chi = 100$ and $\chi = 150$. The results are indistinguishable, a consequence of the very small value of the correlation length, and we have thus not performed calculations for other values of $\chi$.

\subsubsection{Two-step transition}

We recall that for the Pokrovsky-Talapov transition characterised by $\nu_{x+y} = 1, \nu_{x-y} = 1/2 $, one expects to observe $\nu_y  = 1/2$. In contrast, for the Kosterlitz-Thouless transition, due to the exponential divergence of the correlation length in both $x\pm y$ directions, one expects $\nu_{y}$ as well as  $\nu_{y\pm x}$ to diverge from the incommensurate phase at the critical temperature.

\begin{figure}[h!]
\centering
\includegraphics[width = 0.45\textwidth]{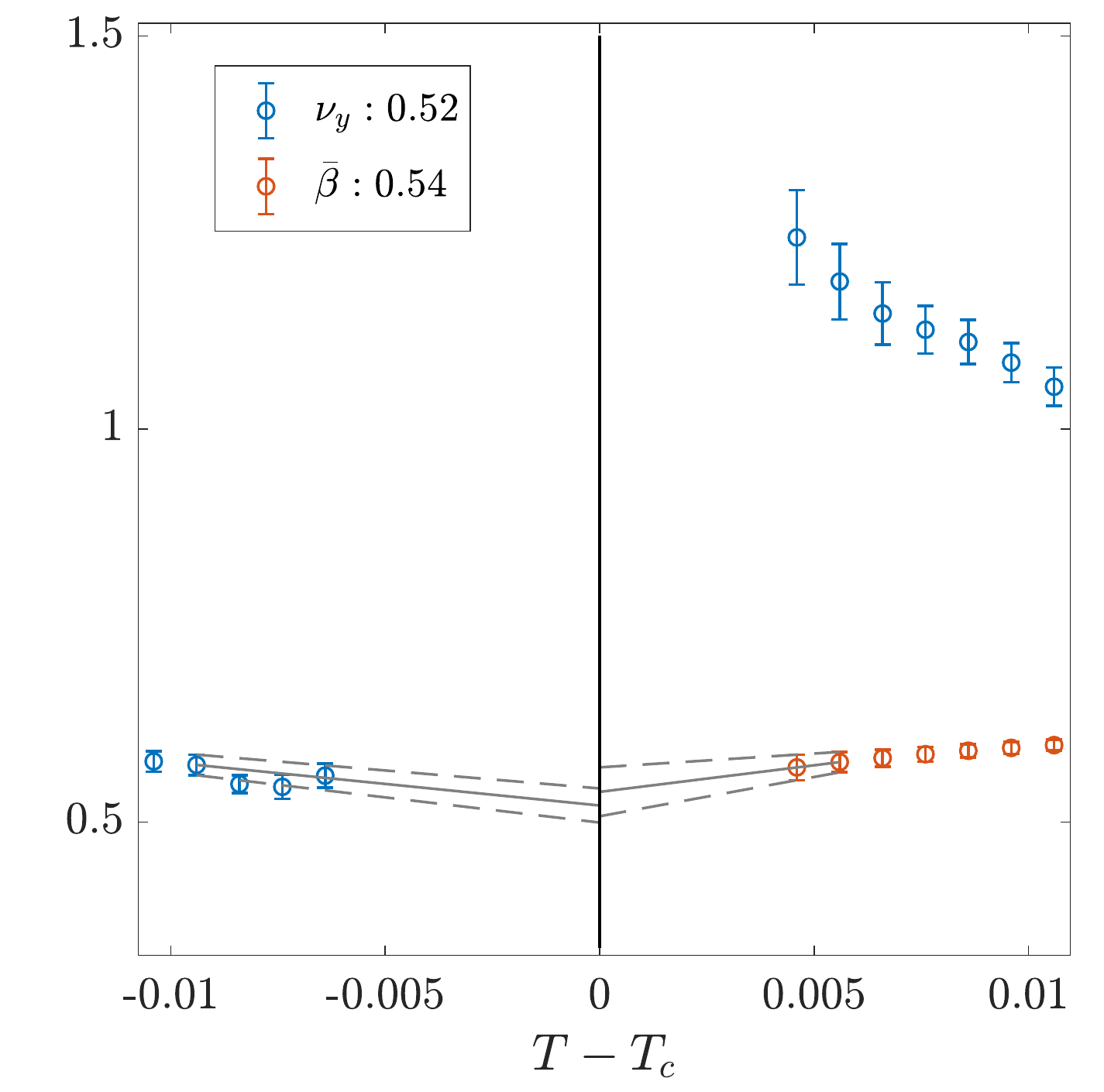}
\caption{Exponents $\bar{\beta}$ and $\nu_y$ as a function of $T-T_c$ at $z = 5$, with a critical temperature $T_c = 0.7714\pm 10^{-4}$. For $\bar{\beta}$, a linear extrapolation on the two last points gives $\bar{\beta} = 0.54\pm 0.03$. For $\nu_y^{LT}$, we have extrapolated on the four last points due to the larger noise and found $\nu_y^{LT} =0.52 \pm 0.02$. The error bars result from a linear fit of the upper and lower values given by the error bars of the effective exponents at a given temperature. One can clearly identify the beginning of the divergence of $\nu_y$ from the incommensurate phase. }
\label{fig:z500exp}
\end{figure}

As expected, in the large $z$ limit we found a two-step transition in agreement with a PT transition from low temperature and a KT transition from high temperature. We discuss the $z = 5$ case in details but similar results were obtained for other activities. By setting the critical temperature $T_{PT}$ such that $\bar{\beta}= \nu_y^{LT}$, we found $\bar{\beta} = 0.54\pm 0.03 $ and $\nu_y^{LT} = 0.52 \pm 0.02$, both in reasonable agreement with the PT universality class exponents. Furthermore, $\nu_y$ from the incommensurate phase diverges, in agreement with a KT transition. The floating phase is however too narrow to determine $T_{KT}$. Indeed, an exponential fit of the correlation length would not be able to distinguish the two critical temperatures. We summarise the results in Fig. \ref{fig:z500exp}.

% dq :  6.746  + 0.5383 % diff de 0.0312
% dq+ : 3.514 + 0.5695
% dq- :  9.978 + 0.5073
% nult: -4.439 + 0.5210 % diff 0.0217
% nult: -3.429 + 0.5427
% nult: -5.449 + 0.4993

\subsubsection{Chiral transition}

\begin{figure}[t!]
\includegraphics[width=.45\textwidth]{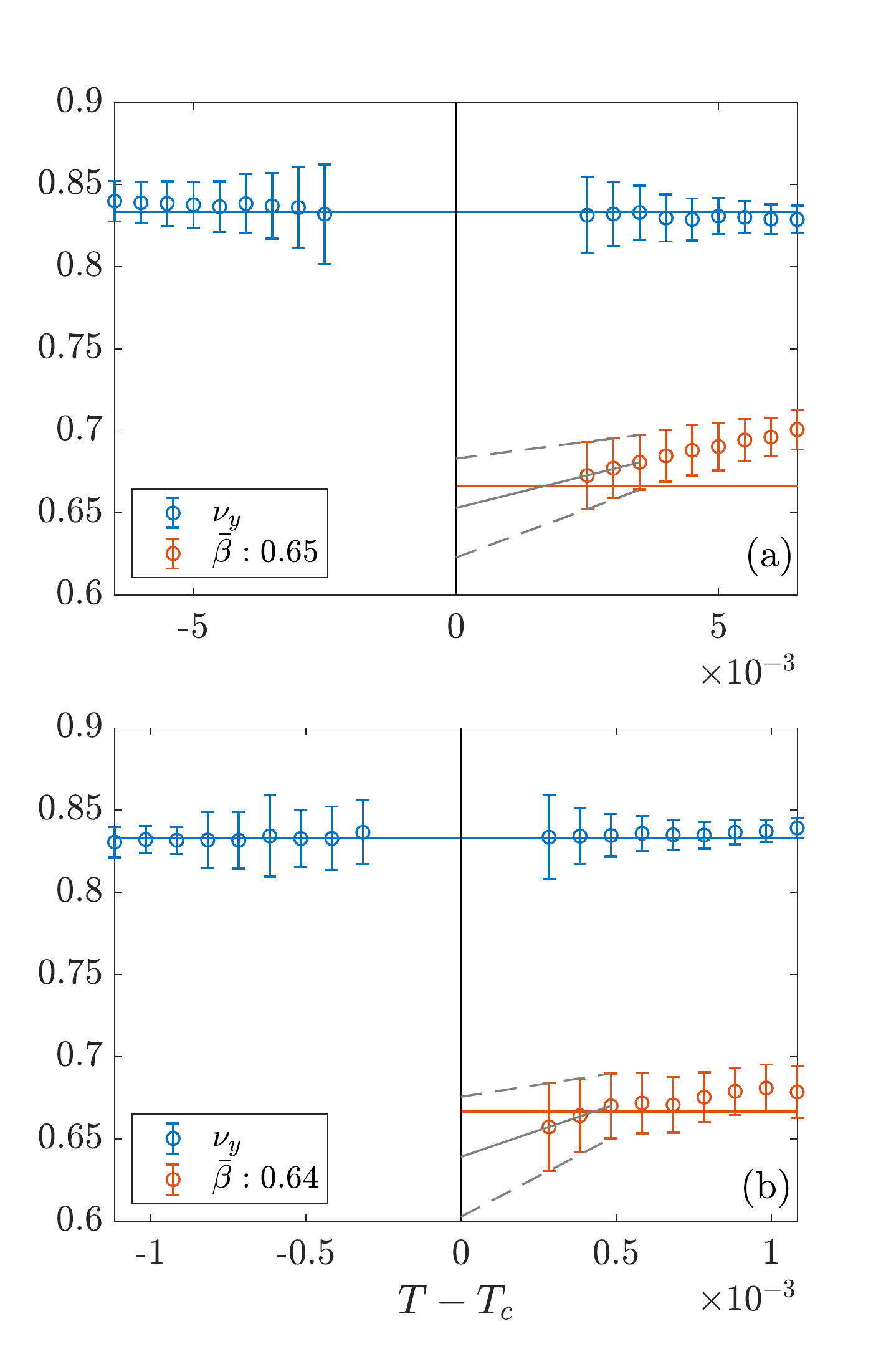}
\caption{Exponents $\bar{\beta}$ and $\nu_y$ as a function of $T-T_c$. The blue and red lines represent the $5/6$ and $2/3$ values. (a) Simulation done to the right of the Potts point at $z=1$ with critical temperature $T_c = 0.7013\pm 5 \cdot 10^{-5}$. (b) Simulations to the left of the Potts point are not done at $z$ or $M$ constant but along a line that crosses the transition at $z_c = 0.57$ with critical temperature $T_c = 0.603916\pm 5\cdot 10^{-6}$. For both activities we found a unique transition characterised by the Potts exponent $\nu$ and $\bar{\beta} = 2/3$. Linear fits on the last three points for $z = 1$ and $z = 0.57$ lead respectively to $\bar{\beta} = 0.65 \pm 0.03 $ and $\bar{\beta} = 0.64 \pm 0.04$.
}
\label{fig:chiraltrans}
\end{figure}

\begin{figure}[t!]
\includegraphics[width=.45\textwidth]{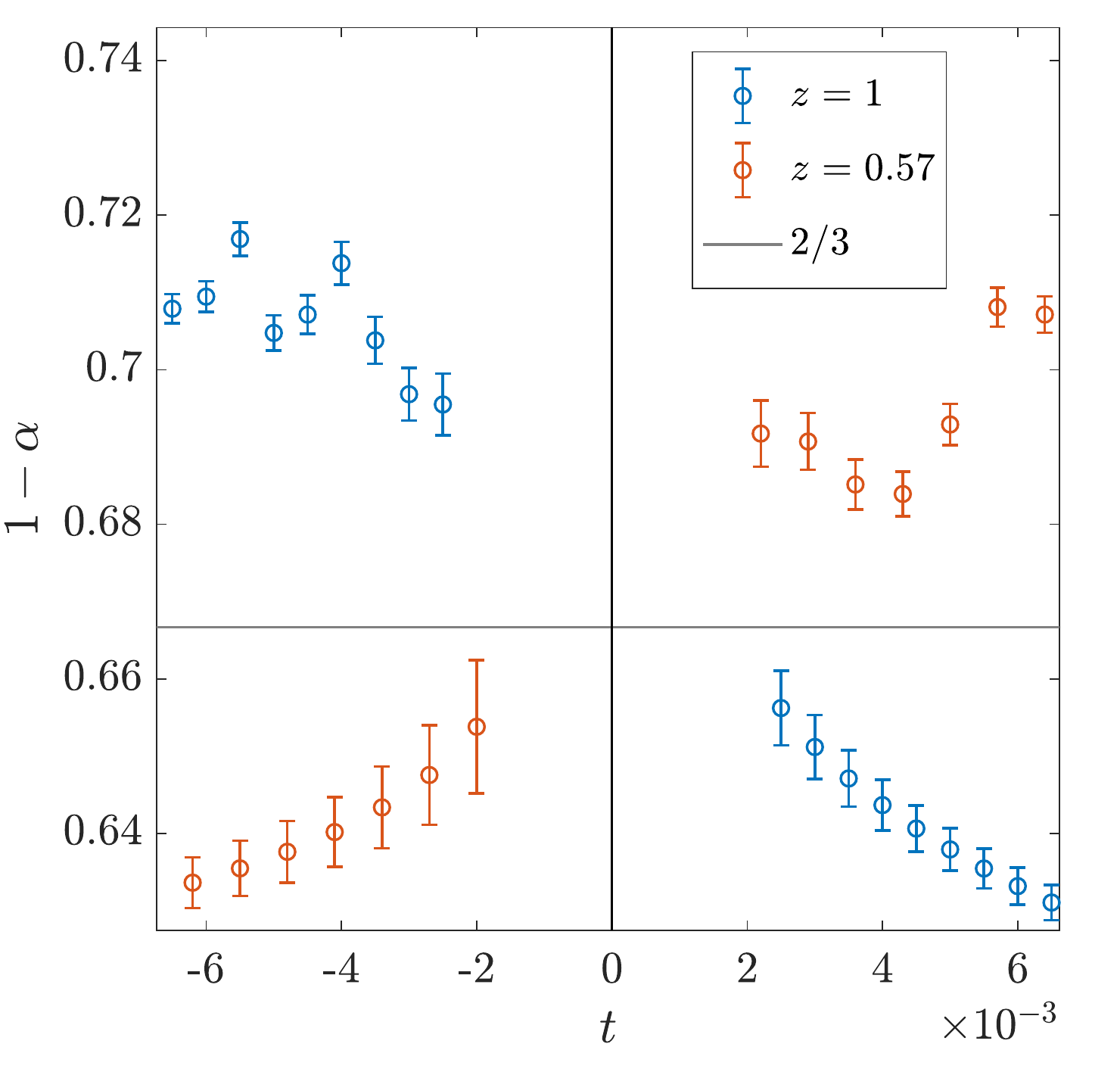}
\caption{Effective exponent of the energy singularity $1-\alpha$ across two points where the transition out of the period-3 phase is direct. For simulations at the critical activity $z_c = 0.57$, $t = z - z_c$,  while for simulations at $z = 1$, $t =T -T_c$. Simulations were performed at $\chi = 200$ and with a convergence criteria of $\delta E = 10^{-9}$. Error bars comes form the estimate of the critical energy. As the simulations were performed at finite bond dimension, we do not extrapolate the exponent but rather show that its behaviour is in agreement with $\alpha = 1/3$. We attribute the noise in the $3\times 1$ phase to the translational symmetry breaking. 
}
\label{fig:Spe}
\end{figure}

We now move to the investigation of the transition close to the Potts point. Over a parameter range covering larger and smaller activities in the vicinity of the Potts point, we found a unique transition, with numerical evidence that it is characterized by $\bar{\beta} = 2/3$ and $\alpha = 1/3$. Results are summarised in Figs. \ref{fig:chiraltrans} and \ref{fig:Spe}. In particular, we found a unique transition with critical exponent $\bar{\beta} = 0.64 \pm 0.04 $ and $\bar{\beta} = 0.65 \pm 0.03$ at activities $z = 0.57$ and $z = 1$ respectively. We notice that in this parameter range we recover an exponent $\nu_y$ in relative agreement with its three-state Potts value $\nu = 5/6$. This is most probably due to a strong crossover effect in the correlation length, as predicted\cite{HuseFisher1984} and observed numerically in systems where chiral transitions are believed to take place\cite{Nyckees2020}.

In order to measure the exponent $\alpha$, it is more accurate to study the singularity of the energy rather than the divergence of the specific heat. This allows one to bypass the error due to the numerical derivative of the energy at the expense of an additional, smaller source of error coming from the estimate of the critical energy. The singularity of the energy, whose exponent is equal to $1-\alpha$, is in good agreement with a critical exponent $\alpha = 1/3$ all along the transition. Hence we found evidence for $\alpha$ to keep its three-state Potts value along the chiral transition. We note that this behaviour has already been observed numerically on models along transitions which are believed to be chiral\cite{Nyckees2020,chepiga_mila_PRL,Chepiga2021}.

The accuracy of the exponent $\bar{\beta}$ over a large range of parameter excludes the possibility of the transition belonging to the three-state Potts universality class. Also, the convergence to a unique limit of $\nu$ from both sides of the transition together with $\bar{\beta}> 1/2$ excludes the two-step transition scenario. We thus conclude that the melting must occur through a unique chiral transition.

% on three point
% z = 057
% 0.6392, 0.64
% 0.6757, 0.68
% 0.6027, 0.60
% z = 100
% 0.6530, 0.65
% 0.6831, 0.68
% 0.6228, 0.62

\subsubsection{Phase diagram}

The phase diagram is presented in Fig. \ref{fig:PhaseDiagram}. We found $\pi$-commensurate and $2\pi/3$-commensurate phases separated by an incommensurate one. The $\pi$-commensurate and incommensurate phases are separated by a disorder line (also sometimes referred to as a Lifshitz transition). Within the $\pi$-commensurate phase, we found an Ising transition whose critical temperature increases upon reducing the activity $z$, consistent with a divergence at the hard-square lattice gas critical activity $z=3.7962$. In contrast, the nature of the $2\pi/3$-commensurate - incommensurate transition depends on the activity. In particular, along the integrable line, we found the transition to belong to the three-state Potts universality class in agreement with Baxter's derivation. We note that the $2\pi/3$-commensurate line leads straight away to the Potts point, in agreement with the chiral operator vanishing at that special point, with $q>2\pi/3$ to the right of the line and $q<2\pi/3$ to its left. On the other hand, in the high activity limit we found a two-step transition separated by a narrow floating phase bounded by a KT transition and a PT transition respectively in the high and low temperature regime. We note that to the left of the Potts point at $z=0.31$ we found a critical exponent $\bar{\beta} = 0.59$, significantly smaller then the believed chiral value $\bar{\beta} = 2/3$ value. This could be explained by the presence of a floating phase to the left of the Potts point as well, in which case we measure a crossover from the PT value $\bar{\beta} = 1/2$. This is in agreement with the numerical results in the hard-boson model\cite{chepiga_mila_PRL} where a floating phase on both sides of the Potts point has already been observed. Finally, in the vicinity of the Potts point, we found evidence for a unique transition characterised by $\bar{\beta} = 2/3$ and $\alpha= 1/3$. Note that we were not able to determine the position of the Lifshitz point with precision. 
%We estimate it to be within $z_L \in [1,5]$. 
This is due partly to the fact that the floating phase is extremely narrow, and to the fact that we do not have access to the correlation and wave-vector along the commensurate direction.

\subsubsection{Discussion}

\begin{figure}[t!]
\centering
\includegraphics[width = .45\textwidth]{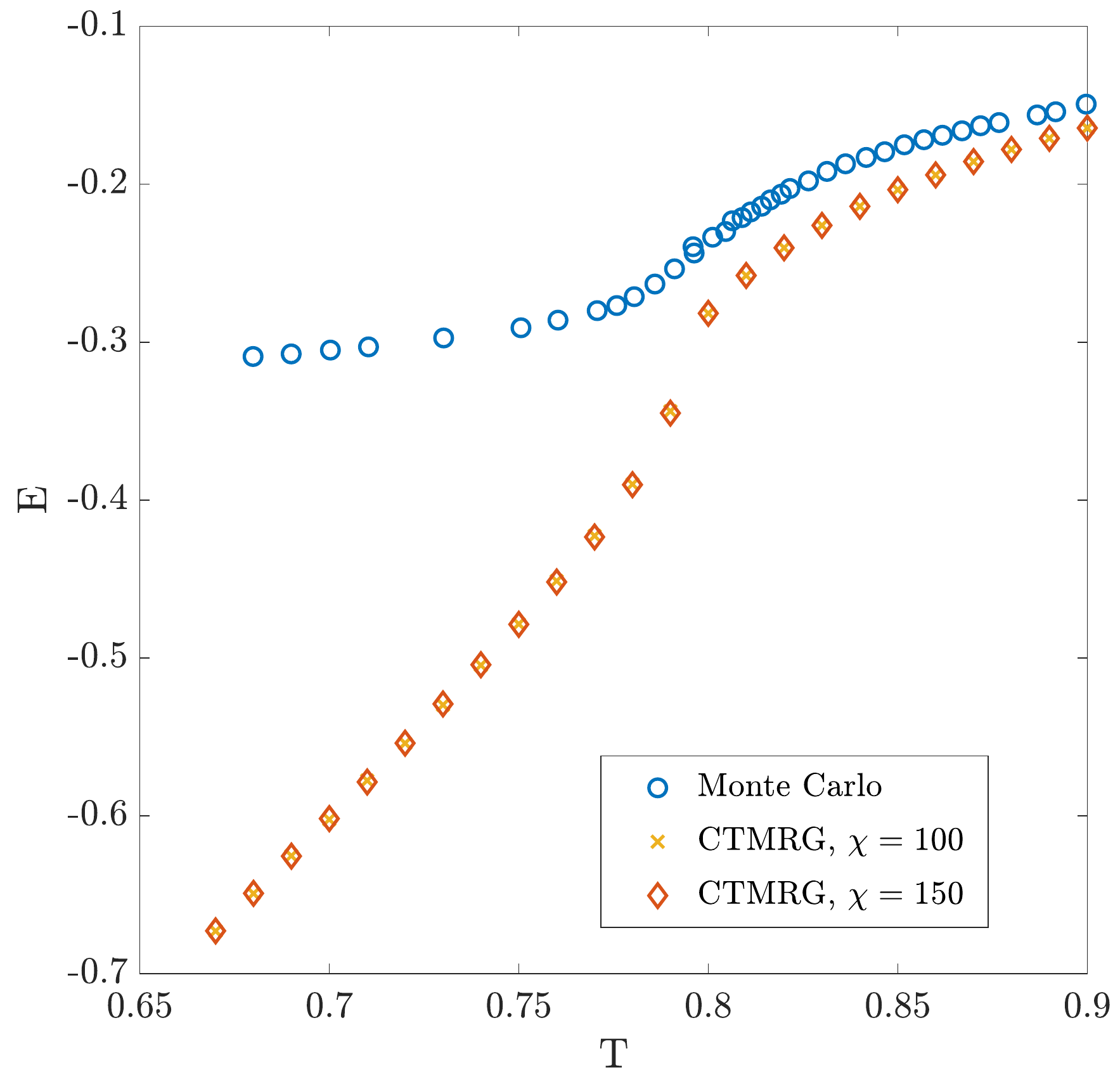} 
\caption{Comparison of Monte Carlo and CTMRG energies as a function of temperature at $z = 2.5$. The CTMRG energy is converged with respect to the bond dimension. The Monte Carlo data come from ref. \cite{bartelt} and were extracted with the use of\cite{Rohatgi2020}.}
\label{fig:EnerTE}
\end{figure}

This cut was already studied by N. C. Bartelt \textit{et al} in the late eighties with Monte-Carlo techniques. In particular they studied the melting of the $3\times1$ ordered phase at different activities to the right of the Potts point and gave evidence for a chiral melting to take place, with a critical exponent $\bar{\beta} \simeq 0.8$, while our results are more consistent with $\bar{\beta} = 2/3$. Quite logically, they have computed correlations along the ${x\pm y}$ directions, something we cannot do, so we are unable to compare our correlation length and wave-vector with theirs. However,  we can still compare energies. They computed the energy along the $z = 2.5$ cut, which we display in Fig. \ref{fig:EnerTE} together with our result. As expected, our critical temperatures, as measured by the change of convexity, are comparable. At high energy, the difference is small, and it is plausible that this is a finite-size effect of Monte Carlo simulations. However, at low temperature, the difference gets too large to  be accounted for by finite size corrections. We believe that this is due to the lost of ergodicity in the Monte-Carlo simulations, a problem that could also explain why the product $(q-q_0)\xi$ remains finite instead of going to to zero in the commensurate phase (Fig. [7] in ref. \cite{bartelt}). 

\subsection{Results for $L = 2$}

\begin{figure}[t!]
\centering
\includegraphics[width = .45\textwidth]{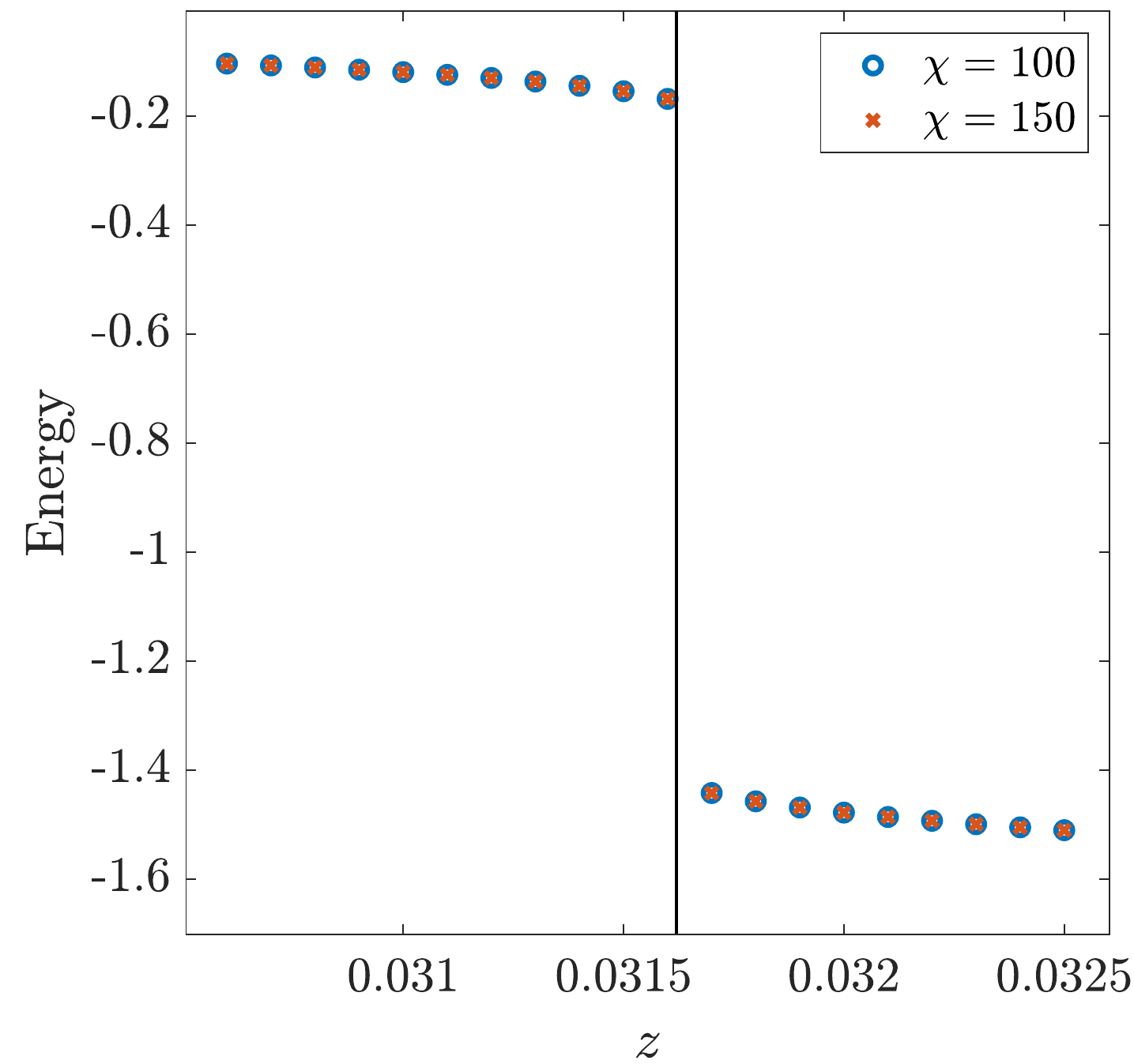} 
\caption{Energy across the transition at $M = 1.5$ and finite bond dimension. The black line represents the integrable point. The energy has converged with respect to the bond dimension, and we clearly observe a discontinuity, in agreement with the presence of a first order transition.}
\label{fig:FirstOrder}
\end{figure}

\begin{figure}[t!]
\centering
\includegraphics[width = .45\textwidth]{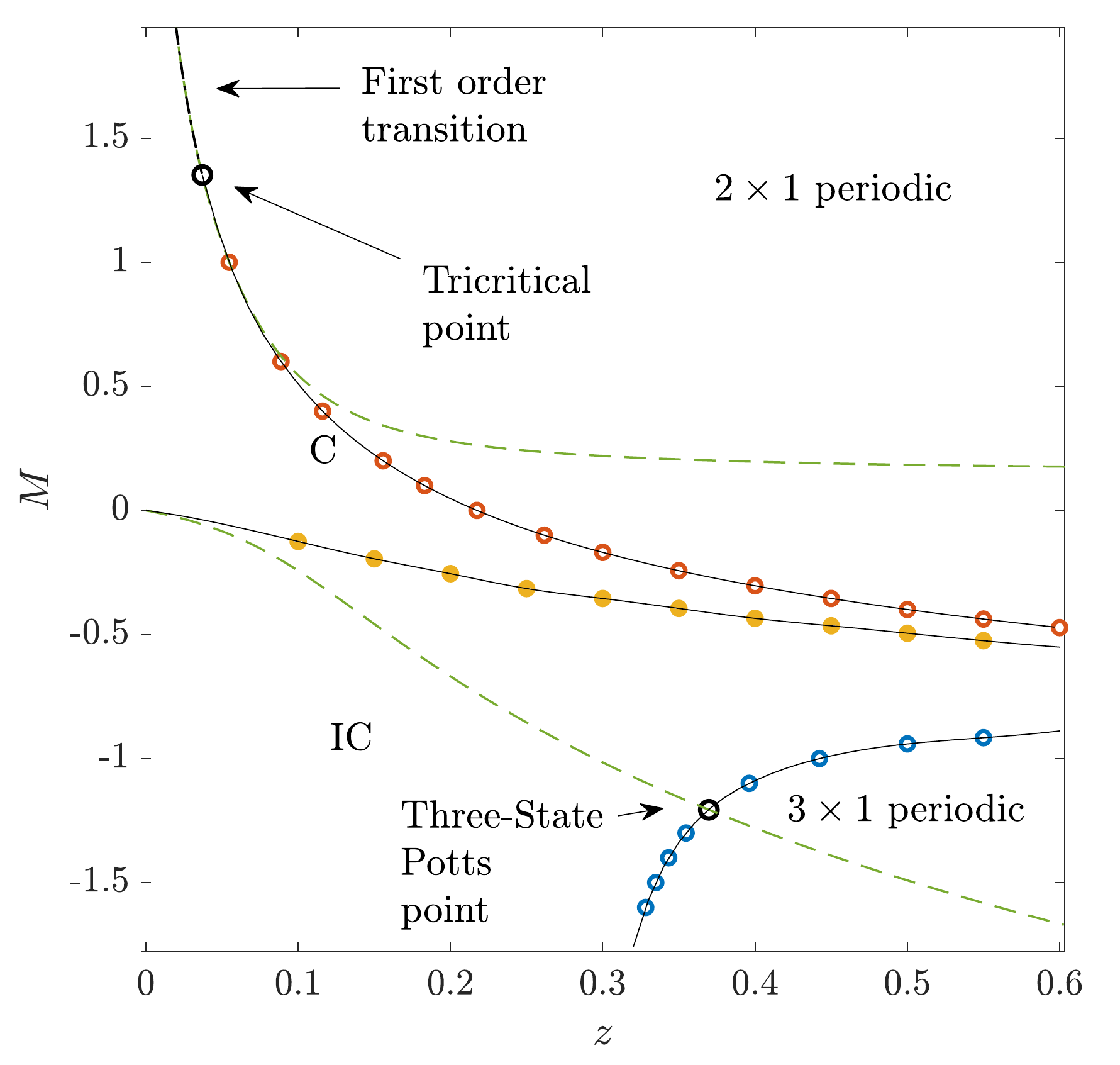} 
\caption{Phase diagram of the model with $L = 2$ where we have plotted  (\protect\tikz \protect\draw[rouge] (0,0) circle (.5ex);)  Ising transition, (\protect\tikz \protect\draw[bleu] (0,0) circle (.5ex);)  chiral / PT transition,(\protect\tikz \protect\draw[jaune] (0,0) circle (.5ex);) order-disorder line, ( \protect\tikz \protect\draw[thick, vert,-] (0,0.1) to (0.2,0.1) ; )  integrable line, black solid lines are guide to the eyes. }
\label{fig:PhaseDiagram2}
\end{figure}

Cuts at fixed $L$ have not been studied before. As explained above, such cuts are interesting because they are closer to the 1D quantum version of the model, but also because they are expected to reveal the full richness of the critical properties of the model, with the presence of both an Ising tricritical point and a three-state Potts point. Indeed, Eq.\ref{eqn:Int2} has two solutions, one for which $M>0$ belonging to the Ising tricritical universality class, and one with $M<0$ that  belongs to the three-state Potts universality class. So we have performed an exhaustive numerical investigation of the $L=2$ cut. The phase diagram of that model as a function of $z$ and $M$
is shown in Fig. \ref{fig:PhaseDiagram2}. All the boundaries have been calculated as for the other cut, and we do not show details for conciseness. The only qualitative difference is the presence of a first-order transition above the Ising tricritical point. Its location is known exactly because it lies in the integrable manifold. Still, for completeness, we have calculated the energy with CTMRG across this cut, as shown in Fig.\ref{fig:FirstOrder}, where we plot the energy as a function of $z$ for $M=1.5$, and it indeeds has an abrupt jump at the transition, as expected for a first-order transition.

As expected, the phase diagram of this cut is very similar to that of the hard-core bosons\cite{fendley,chepiga_mila_PRL}. Interestingly, the two ways of approaching essentially the same problem are not redundant but complementary, and taken together, the results of CTMRG on the classical problem and of density-matrix renormalization group (DMRG) on the quantum problem lead to a solid and consistent picture regarding the nature of the transition out of the period-3 phase. Far enough from the Potts point, and on both sides, the transition takes place through a very narrow floating phase. For that issue, DMRG simulations on very long chains are superior, and floating phase widths smaller than 10$^{-2}$ could be explicitly determined\cite{chepiga_mila_PRL}, something out of range for our CTMRG simulations. Close to the Potts, both approaches lead to the conclusion that the transition is chiral, but the CTMRG simulations consistently find an exponent $\bar \beta \simeq 2/3$ with very good accuracy, while the precise determination of $\bar \beta$ with DMRG is more difficult. Assuming that the hyperscaling relation $\nu_{x+y} + \nu_{x-y} = 2 - \alpha$ holds, and taking for granted that $\alpha$ keeps the value of the Potts point $\alpha=1/3$ along the chiral transition, as suggested by CTMRG, the emerging picture is that of a Potts point surrounded by a chiral transition with exponents $\bar \beta = 2/3$, $\alpha=1/3$, and a dynamical exponent $\nu_{x+y}/\nu_{x-y}=3/2$, with further away a transition through a very narrow floating phase. The only remaining open issue seems to be the precise location of the Lifshitz point that separates the chiral transition from the floating phase.\\

\section{Discussion}

The equivalence between quantum models in dimension D and classical models in dimension D+1 has proven to be extremely powerful to identify the universality class of phase transitions. In particular, 1D quantum models are equivalent to 2D classical models, and if they are conformal, transitions can be expected to belong to one of the minimal models of 2D conformal field theory such as Ising, tricritical Ising, 3-state Potts, etc. If the transition is continuous but non-conformal, as the chiral commensurate-incommensurate transition proposed by Huse and Fisher, the equivalence still holds, but there is no general classification scheme. So to study the 2D classical version of a 1D quantum problem or vice versa might seem to be a pointless exercise. The results reported here show that this can still be rewarding because very sophisticated numerical approaches have been developed for 1D quantum problems and for 2D classical problems, and they appear as complementary to study different aspects of the problem. 

For the hard square model studied here, direct evidence of a floating phase far enough from the Potts point could not be obtained directly, but this information could be obtained with DMRG simulations of the 1D quantum version of the model. When it comes to the exponents of the chiral transition close to the Potts points however, the CTMRG investigation of the classical 2D problem is definitely more accurate. This suggests to see CTMRG not only as a powerful alternative to Monte Carlo for classical problems, but as a complementary tool to study subtle issues in 1D quantum physics. From that point of view, it would be for instance very interesting to use CTMRG to study the 2D classical version of the further-neighbor blockade models\cite{Chepiga2021} that have been proposed as effective models for the higher commensurability phases of chains of Rydberg atoms.

%We have studied two different cuts of the three dimensional phase diagram of the hard-square model with CTMRG, where we observe multiple phases, among which a $\pi$ and 2$\pi/3$-commensurate ones separated by an incommensurate phase. We found within the $\pi$-commensurate phase a transition which seems to either be a first order transition or a continuous one belonging to the Ising universality class. The 2$\pi/3$-commensurate - incommensurate transition is a lot richer and we have identified three different ways for the melting to occur: (i) through a three-state Potts transition along the integrable line in agreement with Baxter computations, (ii) through a two-step process separated by a floating, described by a PT transition from the period three phase and a KT transition in the disordered phase, (iii) through a unique transition for a range of parameter in the vicinity and from both sides of the Potts point along which $\bar{\beta} = 2/3$ and $\alpha = 1/3$. Those exponents cannot describe a three-state Potts transition and then suggest for the transition to be chiral. Such transitions are also believed to take place in the $2$-site blockade model. Thus, in light of the similarities between the hard-square and hard-core Boson model it would be interesting to study the classical equivalent of further $r$-site blockade model as well.

{\it Acknowledgments.} We thank Jeanne Colbois and Zakaria Jouini for useful discussions.
This work has been supported by the Swiss National Science Foundation.
The calculations have been performed using the facilities of the Scientific IT and Application Support Center of EPFL.

\bibliography{bibliography,comments}

\begin{thebibliography}{48}%
\makeatletter
\providecommand \@ifxundefined [1]{%
 \@ifx{#1\undefined}
}%
\providecommand \@ifnum [1]{%
 \ifnum #1\expandafter \@firstoftwo
 \else \expandafter \@secondoftwo
 \fi
}%
\providecommand \@ifx [1]{%
 \ifx #1\expandafter \@firstoftwo
 \else \expandafter \@secondoftwo
 \fi
}%
\providecommand \natexlab [1]{#1}%
\providecommand \enquote  [1]{``#1''}%
\providecommand \bibnamefont  [1]{#1}%
\providecommand \bibfnamefont [1]{#1}%
\providecommand \citenamefont [1]{#1}%
\providecommand \href@noop [0]{\@secondoftwo}%
\providecommand \href [0]{\begingroup \@sanitize@url \@href}%
\providecommand \@href[1]{\@@startlink{#1}\@@href}%
\providecommand \@@href[1]{\endgroup#1\@@endlink}%
\providecommand \@sanitize@url [0]{\catcode `\\12\catcode `\$12\catcode
  `\&12\catcode `\#12\catcode `\^12\catcode `\_12\catcode `\%12\relax}%
\providecommand \@@startlink[1]{}%
\providecommand \@@endlink[0]{}%
\providecommand \url  [0]{\begingroup\@sanitize@url \@url }%
\providecommand \@url [1]{\endgroup\@href {#1}{\urlprefix }}%
\providecommand \urlprefix  [0]{URL }%
\providecommand \Eprint [0]{\href }%
\providecommand \doibase [0]{http://dx.doi.org/}%
\providecommand \selectlanguage [0]{\@gobble}%
\providecommand \bibinfo  [0]{\@secondoftwo}%
\providecommand \bibfield  [0]{\@secondoftwo}%
\providecommand \translation [1]{[#1]}%
\providecommand \BibitemOpen [0]{}%
\providecommand \bibitemStop [0]{}%
\providecommand \bibitemNoStop [0]{.\EOS\space}%
\providecommand \EOS [0]{\spacefactor3000\relax}%
\providecommand \BibitemShut  [1]{\csname bibitem#1\endcsname}%
\let\auto@bib@innerbib\@empty
%</preamble>
\bibitem [{\citenamefont {Bernien}\ \emph {et~al.}(2017)\citenamefont
  {Bernien}, \citenamefont {Schwartz}, \citenamefont {Keesling}, \citenamefont
  {Levine}, \citenamefont {Omran}, \citenamefont {Pichler}, \citenamefont
  {Choi}, \citenamefont {Zibrov}, \citenamefont {Endres}, \citenamefont
  {Greiner}, \citenamefont {Vuletic},\ and\ \citenamefont {Lukin}}]{lukin2017}%
  \BibitemOpen
  \bibfield  {author} {\bibinfo {author} {\bibfnamefont {Hannes}\ \bibnamefont
  {Bernien}}, \bibinfo {author} {\bibfnamefont {Sylvain}\ \bibnamefont
  {Schwartz}}, \bibinfo {author} {\bibfnamefont {Alexander}\ \bibnamefont
  {Keesling}}, \bibinfo {author} {\bibfnamefont {Harry}\ \bibnamefont
  {Levine}}, \bibinfo {author} {\bibfnamefont {Ahmed}\ \bibnamefont {Omran}},
  \bibinfo {author} {\bibfnamefont {Hannes}\ \bibnamefont {Pichler}}, \bibinfo
  {author} {\bibfnamefont {Soonwon}\ \bibnamefont {Choi}}, \bibinfo {author}
  {\bibfnamefont {Alexander~S.}\ \bibnamefont {Zibrov}}, \bibinfo {author}
  {\bibfnamefont {Manuel}\ \bibnamefont {Endres}}, \bibinfo {author}
  {\bibfnamefont {Markus}\ \bibnamefont {Greiner}}, \bibinfo {author}
  {\bibfnamefont {Vladan}\ \bibnamefont {Vuletic}}, \ and\ \bibinfo {author}
  {\bibfnamefont {Mikhail~D.}\ \bibnamefont {Lukin}},\ }\bibfield  {title}
  {\enquote {\bibinfo {title} {Probing many-body dynamics on a 51-atom quantum
  simulator},}\ }\href {http://dx.doi.org/10.1038/nature24622} {\bibfield
  {journal} {\bibinfo  {journal} {Nature}\ }\textbf {\bibinfo {volume} {551}},\
  \bibinfo {pages} {579} (\bibinfo {year} {2017})}\BibitemShut {NoStop}%
\bibitem [{\citenamefont {{Keesling}}\ \emph {et~al.}(2019)\citenamefont
  {{Keesling}}, \citenamefont {{Omran}}, \citenamefont {{Levine}},
  \citenamefont {{Bernien}}, \citenamefont {{Pichler}}, \citenamefont {{Choi}},
  \citenamefont {{Samajdar}}, \citenamefont {{Schwartz}}, \citenamefont
  {{Silvi}}, \citenamefont {{Sachdev}}, \citenamefont {{Zoller}}, \citenamefont
  {{Endres}}, \citenamefont {{Greiner}}, \citenamefont {{Vuleti{\'c}}},
  \citenamefont {{}},\ and\ \citenamefont {{Lukin}}}]{lukin2019}%
  \BibitemOpen
  \bibfield  {author} {\bibinfo {author} {\bibfnamefont {Alexander}\
  \bibnamefont {{Keesling}}}, \bibinfo {author} {\bibfnamefont {Ahmed}\
  \bibnamefont {{Omran}}}, \bibinfo {author} {\bibfnamefont {Harry}\
  \bibnamefont {{Levine}}}, \bibinfo {author} {\bibfnamefont {Hannes}\
  \bibnamefont {{Bernien}}}, \bibinfo {author} {\bibfnamefont {Hannes}\
  \bibnamefont {{Pichler}}}, \bibinfo {author} {\bibfnamefont {Soonwon}\
  \bibnamefont {{Choi}}}, \bibinfo {author} {\bibfnamefont {Rhine}\
  \bibnamefont {{Samajdar}}}, \bibinfo {author} {\bibfnamefont {Sylvain}\
  \bibnamefont {{Schwartz}}}, \bibinfo {author} {\bibfnamefont {Pietro}\
  \bibnamefont {{Silvi}}}, \bibinfo {author} {\bibfnamefont {Subir}\
  \bibnamefont {{Sachdev}}}, \bibinfo {author} {\bibfnamefont {Peter}\
  \bibnamefont {{Zoller}}}, \bibinfo {author} {\bibfnamefont {Manuel}\
  \bibnamefont {{Endres}}}, \bibinfo {author} {\bibfnamefont {Markus}\
  \bibnamefont {{Greiner}}}, \bibinfo {author} {\bibnamefont {{Vuleti{\'c}}}},
  \bibinfo {author} {\bibfnamefont {Vladan}\ \bibnamefont {{}}}, \ and\
  \bibinfo {author} {\bibfnamefont {Mikhail~D.}\ \bibnamefont {{Lukin}}},\
  }\bibfield  {title} {\enquote {\bibinfo {title} {{Quantum Kibble-Zurek
  mechanism and critical dynamics on a programmable Rydberg simulator}},}\
  }\href {\doibase 10.1038/s41586-019-1070-1} {\bibfield  {journal} {\bibinfo
  {journal} {Nature}\ }\textbf {\bibinfo {volume} {568}},\ \bibinfo {pages}
  {207--211} (\bibinfo {year} {2019})}\BibitemShut {NoStop}%
\bibitem [{\citenamefont {Cardy}(1993)}]{cardy}%
  \BibitemOpen
  \bibfield  {author} {\bibinfo {author} {\bibfnamefont {John~L.}\ \bibnamefont
  {Cardy}},\ }\bibfield  {title} {\enquote {\bibinfo {title} {Critical
  exponents of the chiral potts model from conformal field theory},}\ }\href
  {\doibase https://doi.org/10.1016/0550-3213(93)90353-Q} {\bibfield  {journal}
  {\bibinfo  {journal} {Nuclear Physics B}\ }\textbf {\bibinfo {volume}
  {389}},\ \bibinfo {pages} {577 -- 586} (\bibinfo {year} {1993})}\BibitemShut
  {NoStop}%
\bibitem [{\citenamefont {Au-Yang}\ and\ \citenamefont
  {Perk}(1996)}]{AUYANG1996}%
  \BibitemOpen
  \bibfield  {author} {\bibinfo {author} {\bibfnamefont {Helen}\ \bibnamefont
  {Au-Yang}}\ and\ \bibinfo {author} {\bibfnamefont {Jacques~H.H.}\
  \bibnamefont {Perk}},\ }\bibfield  {title} {\enquote {\bibinfo {title} {Phase
  diagram in the generalized chiral clock models},}\ }\href {\doibase
  https://doi.org/10.1016/S0378-4371(96)00058-1} {\bibfield  {journal}
  {\bibinfo  {journal} {Physica A: Statistical Mechanics and its Applications}\
  }\textbf {\bibinfo {volume} {228}},\ \bibinfo {pages} {78 -- 101} (\bibinfo
  {year} {1996})}\BibitemShut {NoStop}%
\bibitem [{\citenamefont {Yeomans}\ and\ \citenamefont
  {Derrida}(1985)}]{yeomans1985}%
  \BibitemOpen
  \bibfield  {author} {\bibinfo {author} {\bibfnamefont {J}~\bibnamefont
  {Yeomans}}\ and\ \bibinfo {author} {\bibfnamefont {B}~\bibnamefont
  {Derrida}},\ }\bibfield  {title} {\enquote {\bibinfo {title} {Bulk and
  interface scaling properties of the chiral clock model},}\ }\href {\doibase
  10.1088/0305-4470/18/12/031} {\bibfield  {journal} {\bibinfo  {journal}
  {Journal of Physics A: Mathematical and General}\ }\textbf {\bibinfo {volume}
  {18}},\ \bibinfo {pages} {2343--2355} (\bibinfo {year} {1985})}\BibitemShut
  {NoStop}%
\bibitem [{\citenamefont {Selke}\ and\ \citenamefont
  {Yeomans}(1982)}]{Selke1982}%
  \BibitemOpen
  \bibfield  {author} {\bibinfo {author} {\bibfnamefont {Walter}\ \bibnamefont
  {Selke}}\ and\ \bibinfo {author} {\bibfnamefont {Julia~M.}\ \bibnamefont
  {Yeomans}},\ }\bibfield  {title} {\enquote {\bibinfo {title} {A monte carlo
  study of the asymmetric clock or chiral potts model in two dimensions},}\
  }\href {\doibase 10.1007/BF01307706} {\bibfield  {journal} {\bibinfo
  {journal} {Zeitschrift f{\"u}r Physik B Condensed Matter}\ }\textbf {\bibinfo
  {volume} {46}},\ \bibinfo {pages} {311--318} (\bibinfo {year}
  {1982})}\BibitemShut {NoStop}%
\bibitem [{\citenamefont {Duxbury}\ \emph {et~al.}(1984)\citenamefont
  {Duxbury}, \citenamefont {Yeomans},\ and\ \citenamefont {Beale}}]{Duxbury}%
  \BibitemOpen
  \bibfield  {author} {\bibinfo {author} {\bibfnamefont {P~M}\ \bibnamefont
  {Duxbury}}, \bibinfo {author} {\bibfnamefont {J}~\bibnamefont {Yeomans}}, \
  and\ \bibinfo {author} {\bibfnamefont {P~D}\ \bibnamefont {Beale}},\
  }\bibfield  {title} {\enquote {\bibinfo {title} {Wavevector scaling and the
  phase diagram of the chiral clock model},}\ }\href
  {http://stacks.iop.org/0305-4470/17/i=4/a=005} {\bibfield  {journal}
  {\bibinfo  {journal} {Journal of Physics A: Mathematical and General}\
  }\textbf {\bibinfo {volume} {17}},\ \bibinfo {pages} {L179} (\bibinfo {year}
  {1984})}\BibitemShut {NoStop}%
\bibitem [{\citenamefont {Sato}\ and\ \citenamefont {Sasaki}(2000)}]{sato}%
  \BibitemOpen
  \bibfield  {author} {\bibinfo {author} {\bibfnamefont {Hiroshi}\ \bibnamefont
  {Sato}}\ and\ \bibinfo {author} {\bibfnamefont {Kazuo}\ \bibnamefont
  {Sasaki}},\ }\bibfield  {title} {\enquote {\bibinfo {title} {Numerical study
  of the two-dimensional three-state chiral clock model by the density matrix
  renormalization group method},}\ }\href {\doibase 10.1143/JPSJ.69.1050}
  {\bibfield  {journal} {\bibinfo  {journal} {Journal of the Physical Society
  of Japan}\ }\textbf {\bibinfo {volume} {69}},\ \bibinfo {pages} {1050--1054}
  (\bibinfo {year} {2000})}\BibitemShut {NoStop}%
\bibitem [{\citenamefont {Houlrik}\ and\ \citenamefont
  {Jensen}(1986)}]{houlrik1986}%
  \BibitemOpen
  \bibfield  {author} {\bibinfo {author} {\bibfnamefont {J.~M.}\ \bibnamefont
  {Houlrik}}\ and\ \bibinfo {author} {\bibfnamefont {S.~J.~Knak}\ \bibnamefont
  {Jensen}},\ }\bibfield  {title} {\enquote {\bibinfo {title} {Phase diagram of
  the three-state chiral clock model studied by monte carlo
  renormalization-group calculations},}\ }\href {\doibase
  10.1103/PhysRevB.34.325} {\bibfield  {journal} {\bibinfo  {journal} {Phys.
  Rev. B}\ }\textbf {\bibinfo {volume} {34}},\ \bibinfo {pages} {325--329}
  (\bibinfo {year} {1986})}\BibitemShut {NoStop}%
\bibitem [{\citenamefont {Au-Yang}\ \emph {et~al.}(1987)\citenamefont
  {Au-Yang}, \citenamefont {McCoy}, \citenamefont {Perk}, \citenamefont
  {Tang},\ and\ \citenamefont {Yan}}]{auyang1987}%
  \BibitemOpen
  \bibfield  {author} {\bibinfo {author} {\bibfnamefont {Helen}\ \bibnamefont
  {Au-Yang}}, \bibinfo {author} {\bibfnamefont {Barry~M.}\ \bibnamefont
  {McCoy}}, \bibinfo {author} {\bibfnamefont {Jacques~H.H.}\ \bibnamefont
  {Perk}}, \bibinfo {author} {\bibfnamefont {Shuang}\ \bibnamefont {Tang}}, \
  and\ \bibinfo {author} {\bibfnamefont {Mu-Lin}\ \bibnamefont {Yan}},\
  }\bibfield  {title} {\enquote {\bibinfo {title} {Commuting transfer matrices
  in the chiral potts models: Solutions of star-triangle equations with
  genus>1},}\ }\href {\doibase https://doi.org/10.1016/0375-9601(87)90065-X}
  {\bibfield  {journal} {\bibinfo  {journal} {Physics Letters A}\ }\textbf
  {\bibinfo {volume} {123}},\ \bibinfo {pages} {219 -- 223} (\bibinfo {year}
  {1987})}\BibitemShut {NoStop}%
\bibitem [{\citenamefont {Baxter}\ \emph {et~al.}(1988)\citenamefont {Baxter},
  \citenamefont {Perk},\ and\ \citenamefont {Au-Yang}}]{baxter1988}%
  \BibitemOpen
  \bibfield  {author} {\bibinfo {author} {\bibfnamefont {R.J.}\ \bibnamefont
  {Baxter}}, \bibinfo {author} {\bibfnamefont {J.H.H.}\ \bibnamefont {Perk}}, \
  and\ \bibinfo {author} {\bibfnamefont {H.}~\bibnamefont {Au-Yang}},\
  }\bibfield  {title} {\enquote {\bibinfo {title} {New solutions of the
  star-triangle relations for the chiral potts model},}\ }\href {\doibase
  https://doi.org/10.1016/0375-9601(88)90896-1} {\bibfield  {journal} {\bibinfo
   {journal} {Physics Letters A}\ }\textbf {\bibinfo {volume} {128}},\ \bibinfo
  {pages} {138 -- 142} (\bibinfo {year} {1988})}\BibitemShut {NoStop}%
\bibitem [{\citenamefont {Schulz}(1980)}]{schulz1980}%
  \BibitemOpen
  \bibfield  {author} {\bibinfo {author} {\bibfnamefont {H.~J.}\ \bibnamefont
  {Schulz}},\ }\bibfield  {title} {\enquote {\bibinfo {title} {Critical
  behavior of commensurate-incommensurate phase transitions in two
  dimensions},}\ }\href {\doibase 10.1103/PhysRevB.22.5274} {\bibfield
  {journal} {\bibinfo  {journal} {Phys. Rev. B}\ }\textbf {\bibinfo {volume}
  {22}},\ \bibinfo {pages} {5274--5277} (\bibinfo {year} {1980})}\BibitemShut
  {NoStop}%
\bibitem [{\citenamefont {Schreiner}\ \emph {et~al.}(1994)\citenamefont
  {Schreiner}, \citenamefont {Jacobi},\ and\ \citenamefont
  {Selke}}]{SelkeExperiment}%
  \BibitemOpen
  \bibfield  {author} {\bibinfo {author} {\bibfnamefont {J.}~\bibnamefont
  {Schreiner}}, \bibinfo {author} {\bibfnamefont {K.}~\bibnamefont {Jacobi}}, \
  and\ \bibinfo {author} {\bibfnamefont {W.}~\bibnamefont {Selke}},\ }\bibfield
   {title} {\enquote {\bibinfo {title} {Experimental evidence for chiral
  melting of the ge(113) and si(113) 3\ifmmode\times\else\texttimes\fi{}1
  surface phases},}\ }\href {\doibase 10.1103/PhysRevB.49.2706} {\bibfield
  {journal} {\bibinfo  {journal} {Phys. Rev. B}\ }\textbf {\bibinfo {volume}
  {49}},\ \bibinfo {pages} {2706--2714} (\bibinfo {year} {1994})}\BibitemShut
  {NoStop}%
\bibitem [{\citenamefont {Howes}(1983)}]{howes1983}%
  \BibitemOpen
  \bibfield  {author} {\bibinfo {author} {\bibfnamefont {Steven~F.}\
  \bibnamefont {Howes}},\ }\bibfield  {title} {\enquote {\bibinfo {title}
  {Commensurate-incommensurate transitions and the lifshitz point in the
  quantum asymmetric clock model},}\ }\href {\doibase 10.1103/PhysRevB.27.1762}
  {\bibfield  {journal} {\bibinfo  {journal} {Phys. Rev. B}\ }\textbf {\bibinfo
  {volume} {27}},\ \bibinfo {pages} {1762--1768} (\bibinfo {year}
  {1983})}\BibitemShut {NoStop}%
\bibitem [{\citenamefont {Fendley}\ \emph {et~al.}(2004)\citenamefont
  {Fendley}, \citenamefont {Sengupta},\ and\ \citenamefont
  {Sachdev}}]{fendley}%
  \BibitemOpen
  \bibfield  {author} {\bibinfo {author} {\bibfnamefont {Paul}\ \bibnamefont
  {Fendley}}, \bibinfo {author} {\bibfnamefont {K.}~\bibnamefont {Sengupta}}, \
  and\ \bibinfo {author} {\bibfnamefont {Subir}\ \bibnamefont {Sachdev}},\
  }\bibfield  {title} {\enquote {\bibinfo {title} {Competing density-wave
  orders in a one-dimensional hard-boson model},}\ }\href {\doibase
  10.1103/PhysRevB.69.075106} {\bibfield  {journal} {\bibinfo  {journal} {Phys.
  Rev. B}\ }\textbf {\bibinfo {volume} {69}},\ \bibinfo {pages} {075106}
  (\bibinfo {year} {2004})}\BibitemShut {NoStop}%
\bibitem [{\citenamefont {Chepiga}\ and\ \citenamefont
  {Mila}(2019)}]{chepiga_mila_PRL}%
  \BibitemOpen
  \bibfield  {author} {\bibinfo {author} {\bibfnamefont {Natalia}\ \bibnamefont
  {Chepiga}}\ and\ \bibinfo {author} {\bibfnamefont {Fr\'ed\'eric}\
  \bibnamefont {Mila}},\ }\bibfield  {title} {\enquote {\bibinfo {title}
  {Floating phase versus chiral transition in a 1d hard-boson model},}\ }\href
  {\doibase 10.1103/PhysRevLett.122.017205} {\bibfield  {journal} {\bibinfo
  {journal} {Phys. Rev. Lett.}\ }\textbf {\bibinfo {volume} {122}},\ \bibinfo
  {pages} {017205} (\bibinfo {year} {2019})}\BibitemShut {NoStop}%
\bibitem [{\citenamefont {Samajdar}\ \emph {et~al.}(2018)\citenamefont
  {Samajdar}, \citenamefont {Choi}, \citenamefont {Pichler}, \citenamefont
  {Lukin},\ and\ \citenamefont {Sachdev}}]{samajdar}%
  \BibitemOpen
  \bibfield  {author} {\bibinfo {author} {\bibfnamefont {Rhine}\ \bibnamefont
  {Samajdar}}, \bibinfo {author} {\bibfnamefont {Soonwon}\ \bibnamefont
  {Choi}}, \bibinfo {author} {\bibfnamefont {Hannes}\ \bibnamefont {Pichler}},
  \bibinfo {author} {\bibfnamefont {Mikhail~D.}\ \bibnamefont {Lukin}}, \ and\
  \bibinfo {author} {\bibfnamefont {Subir}\ \bibnamefont {Sachdev}},\
  }\bibfield  {title} {\enquote {\bibinfo {title} {Numerical study of the
  chiral ${\mathbb{z}}_{3}$ quantum phase transition in one spatial
  dimension},}\ }\href {\doibase 10.1103/PhysRevA.98.023614} {\bibfield
  {journal} {\bibinfo  {journal} {Phys. Rev. A}\ }\textbf {\bibinfo {volume}
  {98}},\ \bibinfo {pages} {023614} (\bibinfo {year} {2018})}\BibitemShut
  {NoStop}%
\bibitem [{\citenamefont {Everts}\ and\ \citenamefont
  {Roder}(1989)}]{Everts_1989}%
  \BibitemOpen
  \bibfield  {author} {\bibinfo {author} {\bibfnamefont {H~U}\ \bibnamefont
  {Everts}}\ and\ \bibinfo {author} {\bibfnamefont {H}~\bibnamefont {Roder}},\
  }\bibfield  {title} {\enquote {\bibinfo {title} {Transfer matrix study of the
  chiral clock model in the hamiltonian limit},}\ }\href {\doibase
  10.1088/0305-4470/22/13/040} {\bibfield  {journal} {\bibinfo  {journal}
  {Journal of Physics A: Mathematical and General}\ }\textbf {\bibinfo {volume}
  {22}},\ \bibinfo {pages} {2475--2494} (\bibinfo {year} {1989})}\BibitemShut
  {NoStop}%
\bibitem [{\citenamefont {Zhuang}\ \emph {et~al.}(2015)\citenamefont {Zhuang},
  \citenamefont {Changlani}, \citenamefont {Tubman},\ and\ \citenamefont
  {Hughes}}]{hughes}%
  \BibitemOpen
  \bibfield  {author} {\bibinfo {author} {\bibfnamefont {Ye}~\bibnamefont
  {Zhuang}}, \bibinfo {author} {\bibfnamefont {Hitesh~J.}\ \bibnamefont
  {Changlani}}, \bibinfo {author} {\bibfnamefont {Norm~M.}\ \bibnamefont
  {Tubman}}, \ and\ \bibinfo {author} {\bibfnamefont {Taylor~L.}\ \bibnamefont
  {Hughes}},\ }\bibfield  {title} {\enquote {\bibinfo {title} {Phase diagram of
  the ${Z}_{3}$ parafermionic chain with chiral interactions},}\ }\href
  {\doibase 10.1103/PhysRevB.92.035154} {\bibfield  {journal} {\bibinfo
  {journal} {Phys. Rev. B}\ }\textbf {\bibinfo {volume} {92}},\ \bibinfo
  {pages} {035154} (\bibinfo {year} {2015})}\BibitemShut {NoStop}%
\bibitem [{\citenamefont {Howes}\ \emph {et~al.}(1983)\citenamefont {Howes},
  \citenamefont {Kadanoff},\ and\ \citenamefont {Nijs}}]{HOWES1983169}%
  \BibitemOpen
  \bibfield  {author} {\bibinfo {author} {\bibfnamefont {Steven}\ \bibnamefont
  {Howes}}, \bibinfo {author} {\bibfnamefont {Leo~P.}\ \bibnamefont
  {Kadanoff}}, \ and\ \bibinfo {author} {\bibfnamefont {Marcel~Den}\
  \bibnamefont {Nijs}},\ }\bibfield  {title} {\enquote {\bibinfo {title}
  {Quantum model for commensurate-incommensurate transitions},}\ }\href
  {\doibase https://doi.org/10.1016/0550-3213(83)90212-2} {\bibfield  {journal}
  {\bibinfo  {journal} {Nuclear Physics B}\ }\textbf {\bibinfo {volume}
  {215}},\ \bibinfo {pages} {169 -- 208} (\bibinfo {year} {1983})}\BibitemShut
  {NoStop}%
\bibitem [{\citenamefont {Whitsitt}\ \emph {et~al.}(2018)\citenamefont
  {Whitsitt}, \citenamefont {Samajdar},\ and\ \citenamefont
  {Sachdev}}]{sachdev_dual}%
  \BibitemOpen
  \bibfield  {author} {\bibinfo {author} {\bibfnamefont {Seth}\ \bibnamefont
  {Whitsitt}}, \bibinfo {author} {\bibfnamefont {Rhine}\ \bibnamefont
  {Samajdar}}, \ and\ \bibinfo {author} {\bibfnamefont {Subir}\ \bibnamefont
  {Sachdev}},\ }\bibfield  {title} {\enquote {\bibinfo {title} {Quantum field
  theory for the chiral clock transition in one spatial dimension},}\ }\href
  {\doibase 10.1103/PhysRevB.98.205118} {\bibfield  {journal} {\bibinfo
  {journal} {Phys. Rev. B}\ }\textbf {\bibinfo {volume} {98}},\ \bibinfo
  {pages} {205118} (\bibinfo {year} {2018})}\BibitemShut {NoStop}%
\bibitem [{\citenamefont {Centen}\ \emph {et~al.}(1982)\citenamefont {Centen},
  \citenamefont {Rittenberg},\ and\ \citenamefont {Marcu}}]{CENTEN1982585}%
  \BibitemOpen
  \bibfield  {author} {\bibinfo {author} {\bibfnamefont {P.}~\bibnamefont
  {Centen}}, \bibinfo {author} {\bibfnamefont {V.}~\bibnamefont {Rittenberg}},
  \ and\ \bibinfo {author} {\bibfnamefont {M.}~\bibnamefont {Marcu}},\
  }\bibfield  {title} {\enquote {\bibinfo {title} {Non-universality in z3
  symmetric spin systems},}\ }\href {\doibase
  https://doi.org/10.1016/0550-3213(82)90079-7} {\bibfield  {journal} {\bibinfo
   {journal} {Nuclear Physics B}\ }\textbf {\bibinfo {volume} {205}},\ \bibinfo
  {pages} {585 -- 600} (\bibinfo {year} {1982})}\BibitemShut {NoStop}%
\bibitem [{\citenamefont {Ostlund}(1981)}]{Ostlund}%
  \BibitemOpen
  \bibfield  {author} {\bibinfo {author} {\bibfnamefont {S.}~\bibnamefont
  {Ostlund}},\ }\bibfield  {title} {\enquote {\bibinfo {title} {Incommensurate
  and commensurate phases in asymmetric clock models},}\ }\href {\doibase
  10.1103/PhysRevB.24.398} {\bibfield  {journal} {\bibinfo  {journal} {Phys.
  Rev. B}\ }\textbf {\bibinfo {volume} {24}},\ \bibinfo {pages} {398--405}
  (\bibinfo {year} {1981})}\BibitemShut {NoStop}%
\bibitem [{\citenamefont {Huse}(1981)}]{Huse1981}%
  \BibitemOpen
  \bibfield  {author} {\bibinfo {author} {\bibfnamefont {David~A.}\
  \bibnamefont {Huse}},\ }\bibfield  {title} {\enquote {\bibinfo {title}
  {Simple three-state model with infinitely many phases},}\ }\href {\doibase
  10.1103/PhysRevB.24.5180} {\bibfield  {journal} {\bibinfo  {journal} {Phys.
  Rev. B}\ }\textbf {\bibinfo {volume} {24}},\ \bibinfo {pages} {5180--5194}
  (\bibinfo {year} {1981})}\BibitemShut {NoStop}%
\bibitem [{\citenamefont {Schulz}(1983)}]{schulz}%
  \BibitemOpen
  \bibfield  {author} {\bibinfo {author} {\bibfnamefont {H.~J.}\ \bibnamefont
  {Schulz}},\ }\bibfield  {title} {\enquote {\bibinfo {title} {Phase
  transitions in monolayers adsorbed on uniaxial substrates},}\ }\href
  {\doibase 10.1103/PhysRevB.28.2746} {\bibfield  {journal} {\bibinfo
  {journal} {Phys. Rev. B}\ }\textbf {\bibinfo {volume} {28}},\ \bibinfo
  {pages} {2746--2749} (\bibinfo {year} {1983})}\BibitemShut {NoStop}%
\bibitem [{\citenamefont {Huse}\ and\ \citenamefont
  {Fisher}(1984)}]{HuseFisher1984}%
  \BibitemOpen
  \bibfield  {author} {\bibinfo {author} {\bibfnamefont {David~A.}\
  \bibnamefont {Huse}}\ and\ \bibinfo {author} {\bibfnamefont {Michael~E.}\
  \bibnamefont {Fisher}},\ }\bibfield  {title} {\enquote {\bibinfo {title}
  {Commensurate melting, domain walls, and dislocations},}\ }\href {\doibase
  10.1103/PhysRevB.29.239} {\bibfield  {journal} {\bibinfo  {journal} {Phys.
  Rev. B}\ }\textbf {\bibinfo {volume} {29}},\ \bibinfo {pages} {239--270}
  (\bibinfo {year} {1984})}\BibitemShut {NoStop}%
\bibitem [{\citenamefont {Huse}\ and\ \citenamefont
  {Fisher}(1982)}]{HuseFisher1982}%
  \BibitemOpen
  \bibfield  {author} {\bibinfo {author} {\bibfnamefont {David~A.}\
  \bibnamefont {Huse}}\ and\ \bibinfo {author} {\bibfnamefont {Michael~E.}\
  \bibnamefont {Fisher}},\ }\bibfield  {title} {\enquote {\bibinfo {title}
  {Domain walls and the melting of commensurate surface phases},}\ }\href
  {\doibase 10.1103/PhysRevLett.49.793} {\bibfield  {journal} {\bibinfo
  {journal} {Phys. Rev. Lett.}\ }\textbf {\bibinfo {volume} {49}},\ \bibinfo
  {pages} {793--796} (\bibinfo {year} {1982})}\BibitemShut {NoStop}%
\bibitem [{\citenamefont {Pokrovsky}\ and\ \citenamefont
  {Talapov}(1979)}]{Pokrovsky_Talapov}%
  \BibitemOpen
  \bibfield  {author} {\bibinfo {author} {\bibfnamefont {V.~L.}\ \bibnamefont
  {Pokrovsky}}\ and\ \bibinfo {author} {\bibfnamefont {A.~L.}\ \bibnamefont
  {Talapov}},\ }\bibfield  {title} {\enquote {\bibinfo {title} {Ground state,
  spectrum, and phase diagram of two-dimensional incommensurate crystals},}\
  }\href {\doibase 10.1103/PhysRevLett.42.65} {\bibfield  {journal} {\bibinfo
  {journal} {Phys. Rev. Lett.}\ }\textbf {\bibinfo {volume} {42}},\ \bibinfo
  {pages} {65--67} (\bibinfo {year} {1979})}\BibitemShut {NoStop}%
\bibitem [{\citenamefont {Kosterlitz}\ and\ \citenamefont
  {Thouless}(1973)}]{Kosterlitz_Thouless_1973}%
  \BibitemOpen
  \bibfield  {author} {\bibinfo {author} {\bibfnamefont {J~M}\ \bibnamefont
  {Kosterlitz}}\ and\ \bibinfo {author} {\bibfnamefont {D~J}\ \bibnamefont
  {Thouless}},\ }\bibfield  {title} {\enquote {\bibinfo {title} {Ordering,
  metastability and phase transitions in two-dimensional systems},}\ }\href
  {http://stacks.iop.org/0022-3719/6/i=7/a=010} {\bibfield  {journal} {\bibinfo
   {journal} {Journal of Physics C: Solid State Physics}\ }\textbf {\bibinfo
  {volume} {6}},\ \bibinfo {pages} {1181} (\bibinfo {year} {1973})}\BibitemShut
  {NoStop}%
\bibitem [{\citenamefont {den Nijs}(1988)}]{Den_Nijs}%
  \BibitemOpen
  \bibfield  {author} {\bibinfo {author} {\bibfnamefont {Marcel}\ \bibnamefont
  {den Nijs}},\ }\bibfield  {title} {\enquote {\bibinfo {title} {The domain
  wall theory of two-dimensional commensurate-incommensurate phase
  transitions},}\ }\href@noop {} {\bibfield  {journal} {\bibinfo  {journal}
  {Phase Transitions and Critical Phenomena}\ }\textbf {\bibinfo {volume}
  {12}},\ \bibinfo {pages} {219} (\bibinfo {year} {1988})}\BibitemShut
  {NoStop}%
\bibitem [{\citenamefont {Baxter}(1980)}]{baxter1980}%
  \BibitemOpen
  \bibfield  {author} {\bibinfo {author} {\bibfnamefont {R~J}\ \bibnamefont
  {Baxter}},\ }\bibfield  {title} {\enquote {\bibinfo {title} {Hard hexagons:
  exact solution},}\ }\href {\doibase 10.1088/0305-4470/13/3/007} {\bibfield
  {journal} {\bibinfo  {journal} {Journal of Physics A: Mathematical and
  General}\ }\textbf {\bibinfo {volume} {13}},\ \bibinfo {pages} {L61--L70}
  (\bibinfo {year} {1980})}\BibitemShut {NoStop}%
\bibitem [{\citenamefont {Baxter}(1981)}]{Baxter1981}%
  \BibitemOpen
  \bibfield  {author} {\bibinfo {author} {\bibfnamefont {R.~J.}\ \bibnamefont
  {Baxter}},\ }\bibfield  {title} {\enquote {\bibinfo {title} {Rogers-ramanujan
  identities in the hard hexagon model},}\ }\href {\doibase 10.1007/BF01011427}
  {\bibfield  {journal} {\bibinfo  {journal} {Journal of Statistical Physics}\
  }\textbf {\bibinfo {volume} {26}},\ \bibinfo {pages} {427--452} (\bibinfo
  {year} {1981})}\BibitemShut {NoStop}%
\bibitem [{\citenamefont {Huse}(1983)}]{Huse_1983}%
  \BibitemOpen
  \bibfield  {author} {\bibinfo {author} {\bibfnamefont {D~A}\ \bibnamefont
  {Huse}},\ }\bibfield  {title} {\enquote {\bibinfo {title} {Multicritical
  scaling in baxter{\textquotesingle}s hard square lattice gas},}\ }\href
  {\doibase 10.1088/0305-4470/16/18/035} {\bibfield  {journal} {\bibinfo
  {journal} {Journal of Physics A: Mathematical and General}\ }\textbf
  {\bibinfo {volume} {16}},\ \bibinfo {pages} {4357--4368} (\bibinfo {year}
  {1983})}\BibitemShut {NoStop}%
\bibitem [{\citenamefont {Nyckees}\ \emph {et~al.}(2021)\citenamefont
  {Nyckees}, \citenamefont {Colbois},\ and\ \citenamefont
  {Mila}}]{Nyckees2020}%
  \BibitemOpen
  \bibfield  {author} {\bibinfo {author} {\bibfnamefont {Samuel}\ \bibnamefont
  {Nyckees}}, \bibinfo {author} {\bibfnamefont {Jeanne}\ \bibnamefont
  {Colbois}}, \ and\ \bibinfo {author} {\bibfnamefont {Frédéric}\
  \bibnamefont {Mila}},\ }\bibfield  {title} {\enquote {\bibinfo {title}
  {Identifying the huse-fisher universality class of the three-state chiral
  potts model},}\ }\href {\doibase
  https://doi.org/10.1016/j.nuclphysb.2021.115365} {\bibfield  {journal}
  {\bibinfo  {journal} {Nuclear Physics B}\ }\textbf {\bibinfo {volume}
  {965}},\ \bibinfo {pages} {115365} (\bibinfo {year} {2021})}\BibitemShut
  {NoStop}%
\bibitem [{\citenamefont {Abernathy}\ \emph {et~al.}(1994)\citenamefont
  {Abernathy}, \citenamefont {Song}, \citenamefont {Blum}, \citenamefont
  {Birgeneau},\ and\ \citenamefont {Mochrie}}]{abernathy}%
  \BibitemOpen
  \bibfield  {author} {\bibinfo {author} {\bibfnamefont {D.~L.}\ \bibnamefont
  {Abernathy}}, \bibinfo {author} {\bibfnamefont {S.}~\bibnamefont {Song}},
  \bibinfo {author} {\bibfnamefont {K.~I.}\ \bibnamefont {Blum}}, \bibinfo
  {author} {\bibfnamefont {R.~J.}\ \bibnamefont {Birgeneau}}, \ and\ \bibinfo
  {author} {\bibfnamefont {S.~G.~J.}\ \bibnamefont {Mochrie}},\ }\bibfield
  {title} {\enquote {\bibinfo {title} {Chiral melting of the si(113)
  (3\ifmmode\times\else\texttimes\fi{}1) reconstruction},}\ }\href {\doibase
  10.1103/PhysRevB.49.2691} {\bibfield  {journal} {\bibinfo  {journal} {Phys.
  Rev. B}\ }\textbf {\bibinfo {volume} {49}},\ \bibinfo {pages} {2691--2705}
  (\bibinfo {year} {1994})}\BibitemShut {NoStop}%
\bibitem [{\citenamefont {Nishino}\ and\ \citenamefont
  {Okunishi}(1996)}]{nishino}%
  \BibitemOpen
  \bibfield  {author} {\bibinfo {author} {\bibfnamefont {T.}~\bibnamefont
  {Nishino}}\ and\ \bibinfo {author} {\bibfnamefont {K.}~\bibnamefont
  {Okunishi}},\ }\bibfield  {title} {\enquote {\bibinfo {title} {Corner
  transfer matrix renormalization group method},}\ }\href {\doibase
  10.1143/JPSJ.65.891} {\bibfield  {journal} {\bibinfo  {journal} {J. Phys.
  Soc. Jpn.}\ }\textbf {\bibinfo {volume} {65}},\ \bibinfo {pages} {891}
  (\bibinfo {year} {1996})}\BibitemShut {NoStop}%
\bibitem [{\citenamefont {Nyckees}\ and\ \citenamefont
  {Mila}(2022)}]{Nyckees2022}%
  \BibitemOpen
  \bibfield  {author} {\bibinfo {author} {\bibfnamefont {Samuel}\ \bibnamefont
  {Nyckees}}\ and\ \bibinfo {author} {\bibfnamefont {Fr\'ed\'eric}\
  \bibnamefont {Mila}},\ }\bibfield  {title} {\enquote {\bibinfo {title}
  {Commensurate-incommensurate transition in the chiral ashkin-teller model},}\
  }\href {\doibase 10.1103/PhysRevResearch.4.013093} {\bibfield  {journal}
  {\bibinfo  {journal} {Phys. Rev. Research}\ }\textbf {\bibinfo {volume}
  {4}},\ \bibinfo {pages} {013093} (\bibinfo {year} {2022})}\BibitemShut
  {NoStop}%
\bibitem [{\citenamefont {Bartelt}\ \emph {et~al.}(1987)\citenamefont
  {Bartelt}, \citenamefont {Einstein},\ and\ \citenamefont
  {Roelofs}}]{bartelt}%
  \BibitemOpen
  \bibfield  {author} {\bibinfo {author} {\bibfnamefont {N.~C.}\ \bibnamefont
  {Bartelt}}, \bibinfo {author} {\bibfnamefont {T.~L.}\ \bibnamefont
  {Einstein}}, \ and\ \bibinfo {author} {\bibfnamefont {L.~D.}\ \bibnamefont
  {Roelofs}},\ }\bibfield  {title} {\enquote {\bibinfo {title} {Structure
  factors associated with the melting of a (31) ordered phase on a
  centered-rectangular lattice gas: Effective scaling in a three-state
  chiral-clock-like model},}\ }\href {\doibase 10.1103/PhysRevB.35.4812}
  {\bibfield  {journal} {\bibinfo  {journal} {Phys. Rev. B}\ }\textbf {\bibinfo
  {volume} {35}},\ \bibinfo {pages} {4812--4818} (\bibinfo {year}
  {1987})}\BibitemShut {NoStop}%
\bibitem [{\citenamefont {Huse}(1982)}]{HuseHardSquare}%
  \BibitemOpen
  \bibfield  {author} {\bibinfo {author} {\bibfnamefont {David~A.}\
  \bibnamefont {Huse}},\ }\bibfield  {title} {\enquote {\bibinfo {title}
  {Tricriticality of interacting hard squares: Some exact results},}\ }\href
  {\doibase 10.1103/PhysRevLett.49.1121} {\bibfield  {journal} {\bibinfo
  {journal} {Phys. Rev. Lett.}\ }\textbf {\bibinfo {volume} {49}},\ \bibinfo
  {pages} {1121--1124} (\bibinfo {year} {1982})}\BibitemShut {NoStop}%
\bibitem [{\citenamefont {Baxter}\ \emph {et~al.}(1980)\citenamefont {Baxter},
  \citenamefont {Enting},\ and\ \citenamefont {Tsang}}]{Baxter1980LG}%
  \BibitemOpen
  \bibfield  {author} {\bibinfo {author} {\bibfnamefont {R.~J.}\ \bibnamefont
  {Baxter}}, \bibinfo {author} {\bibfnamefont {I.~G.}\ \bibnamefont {Enting}},
  \ and\ \bibinfo {author} {\bibfnamefont {S.~K.}\ \bibnamefont {Tsang}},\
  }\bibfield  {title} {\enquote {\bibinfo {title} {Hard-square lattice gas},}\
  }\href {\doibase 10.1007/BF01012867} {\bibfield  {journal} {\bibinfo
  {journal} {Journal of Statistical Physics}\ }\textbf {\bibinfo {volume}
  {22}},\ \bibinfo {pages} {465--489} (\bibinfo {year} {1980})}\BibitemShut
  {NoStop}%
\bibitem [{\citenamefont {Guo}\ and\ \citenamefont {Blöte}(2002)}]{Guo2002}%
  \BibitemOpen
  \bibfield  {author} {\bibinfo {author} {\bibfnamefont {Wenan}\ \bibnamefont
  {Guo}}\ and\ \bibinfo {author} {\bibfnamefont {Henk}\ \bibnamefont
  {Blöte}},\ }\bibfield  {title} {\enquote {\bibinfo {title} {Finite-size
  analysis of the hard-square lattice gas},}\ }\href {\doibase
  10.1103/PhysRevE.66.046140} {\bibfield  {journal} {\bibinfo  {journal}
  {Physical review. E, Statistical, nonlinear, and soft matter physics}\
  }\textbf {\bibinfo {volume} {66}},\ \bibinfo {pages} {046140} (\bibinfo
  {year} {2002})}\BibitemShut {NoStop}%
\bibitem [{\citenamefont {Schollw\"ock}\ \emph {et~al.}(1996)\citenamefont
  {Schollw\"ock}, \citenamefont {Jolic\oe{}ur},\ and\ \citenamefont
  {Garel}}]{schollwoeck_bilbiq}%
  \BibitemOpen
  \bibfield  {author} {\bibinfo {author} {\bibfnamefont {U.}~\bibnamefont
  {Schollw\"ock}}, \bibinfo {author} {\bibfnamefont {Th.}\ \bibnamefont
  {Jolic\oe{}ur}}, \ and\ \bibinfo {author} {\bibfnamefont {T.}~\bibnamefont
  {Garel}},\ }\bibfield  {title} {\enquote {\bibinfo {title} {Onset of
  incommensurability at the valence-bond-solid point in the s=1 quantum spin
  chain},}\ }\href {\doibase 10.1103/PhysRevB.53.3304} {\bibfield  {journal}
  {\bibinfo  {journal} {Phys. Rev. B}\ }\textbf {\bibinfo {volume} {53}},\
  \bibinfo {pages} {3304--3311} (\bibinfo {year} {1996})}\BibitemShut {NoStop}%
\bibitem [{\citenamefont {Chepiga}\ and\ \citenamefont
  {Mila}(2021)}]{Chepiga2021}%
  \BibitemOpen
  \bibfield  {author} {\bibinfo {author} {\bibfnamefont {Natalia}\ \bibnamefont
  {Chepiga}}\ and\ \bibinfo {author} {\bibfnamefont {Fr{\'e}d{\'e}ric}\
  \bibnamefont {Mila}},\ }\bibfield  {title} {\enquote {\bibinfo {title}
  {Kibble-zurek exponent and chiral transition of the period-4 phase of rydberg
  chains},}\ }\href {\doibase 10.1038/s41467-020-20641-y} {\bibfield  {journal}
  {\bibinfo  {journal} {Nature Communications}\ }\textbf {\bibinfo {volume}
  {12}},\ \bibinfo {pages} {414} (\bibinfo {year} {2021})}\BibitemShut
  {NoStop}%
\bibitem [{\citenamefont {Rohatgi}(2021)}]{Rohatgi2020}%
  \BibitemOpen
  \bibfield  {author} {\bibinfo {author} {\bibfnamefont {Ankit}\ \bibnamefont
  {Rohatgi}},\ }\href {https://automeris.io/WebPlotDigitizer} {\enquote
  {\bibinfo {title} {Webplotdigitizer: Version 4.5},}\ } (\bibinfo {year}
  {2021})\BibitemShut {NoStop}%
\bibitem [{\citenamefont {Baxter}(1982)}]{BaxterBook}%
  \BibitemOpen
  \bibfield  {author} {\bibinfo {author} {\bibfnamefont {R.~J.}\ \bibnamefont
  {Baxter}},\ }\href@noop {} {\emph {\bibinfo {title} {Exactly solved models in
  statistical mechanics}}}\ (\bibinfo {year} {1982})\BibitemShut {NoStop}%
\bibitem [{\citenamefont {White}(1992)}]{dmrg1}%
  \BibitemOpen
  \bibfield  {author} {\bibinfo {author} {\bibfnamefont {Steven~R.}\
  \bibnamefont {White}},\ }\bibfield  {title} {\enquote {\bibinfo {title}
  {Density matrix formulation for quantum renormalization groups},}\ }\href
  {\doibase 10.1103/PhysRevLett.69.2863} {\bibfield  {journal} {\bibinfo
  {journal} {Phys. Rev. Lett.}\ }\textbf {\bibinfo {volume} {69}},\ \bibinfo
  {pages} {2863--2866} (\bibinfo {year} {1992})}\BibitemShut {NoStop}%
\bibitem [{\citenamefont {Or\'us}\ and\ \citenamefont {Vidal}(2009)}]{orus}%
  \BibitemOpen
  \bibfield  {author} {\bibinfo {author} {\bibfnamefont {R.}~\bibnamefont
  {Or\'us}}\ and\ \bibinfo {author} {\bibfnamefont {G.}~\bibnamefont {Vidal}},\
  }\bibfield  {title} {\enquote {\bibinfo {title} {Simulation of
  two-dimensional quantum systems on an infinite lattice revisited: Corner
  transfer matrix for tensor contraction},}\ }\href {\doibase
  10.1103/PhysRevB.80.094403} {\bibfield  {journal} {\bibinfo  {journal} {Phys.
  Rev. B}\ }\textbf {\bibinfo {volume} {80}},\ \bibinfo {pages} {094403}
  (\bibinfo {year} {2009})}\BibitemShut {NoStop}%
\bibitem [{\citenamefont {Rams}\ \emph {et~al.}(2018)\citenamefont {Rams},
  \citenamefont {Czarnik},\ and\ \citenamefont {Cincio}}]{czarnik2018}%
  \BibitemOpen
  \bibfield  {author} {\bibinfo {author} {\bibfnamefont {Marek~M.}\
  \bibnamefont {Rams}}, \bibinfo {author} {\bibfnamefont {Piotr}\ \bibnamefont
  {Czarnik}}, \ and\ \bibinfo {author} {\bibfnamefont {Lukasz}\ \bibnamefont
  {Cincio}},\ }\bibfield  {title} {\enquote {\bibinfo {title} {Precise
  extrapolation of the correlation function asymptotics in uniform tensor
  network states with application to the bose-hubbard and xxz models},}\ }\href
  {\doibase 10.1103/PhysRevX.8.041033} {\bibfield  {journal} {\bibinfo
  {journal} {Phys. Rev. X}\ }\textbf {\bibinfo {volume} {8}},\ \bibinfo {pages}
  {041033} (\bibinfo {year} {2018})}\BibitemShut {NoStop}%
\end{thebibliography}%

\appendix

%\section{Methodology}

%In this section we discuss the algorithm and the methodology we used to conduct the study. The same methodology was already used in previous studies in which more details are available\cite{nyckees2020}.

\section{CTMRG}

\begin{figure}[t!]
\centering
\includegraphics[width = .45\textwidth]{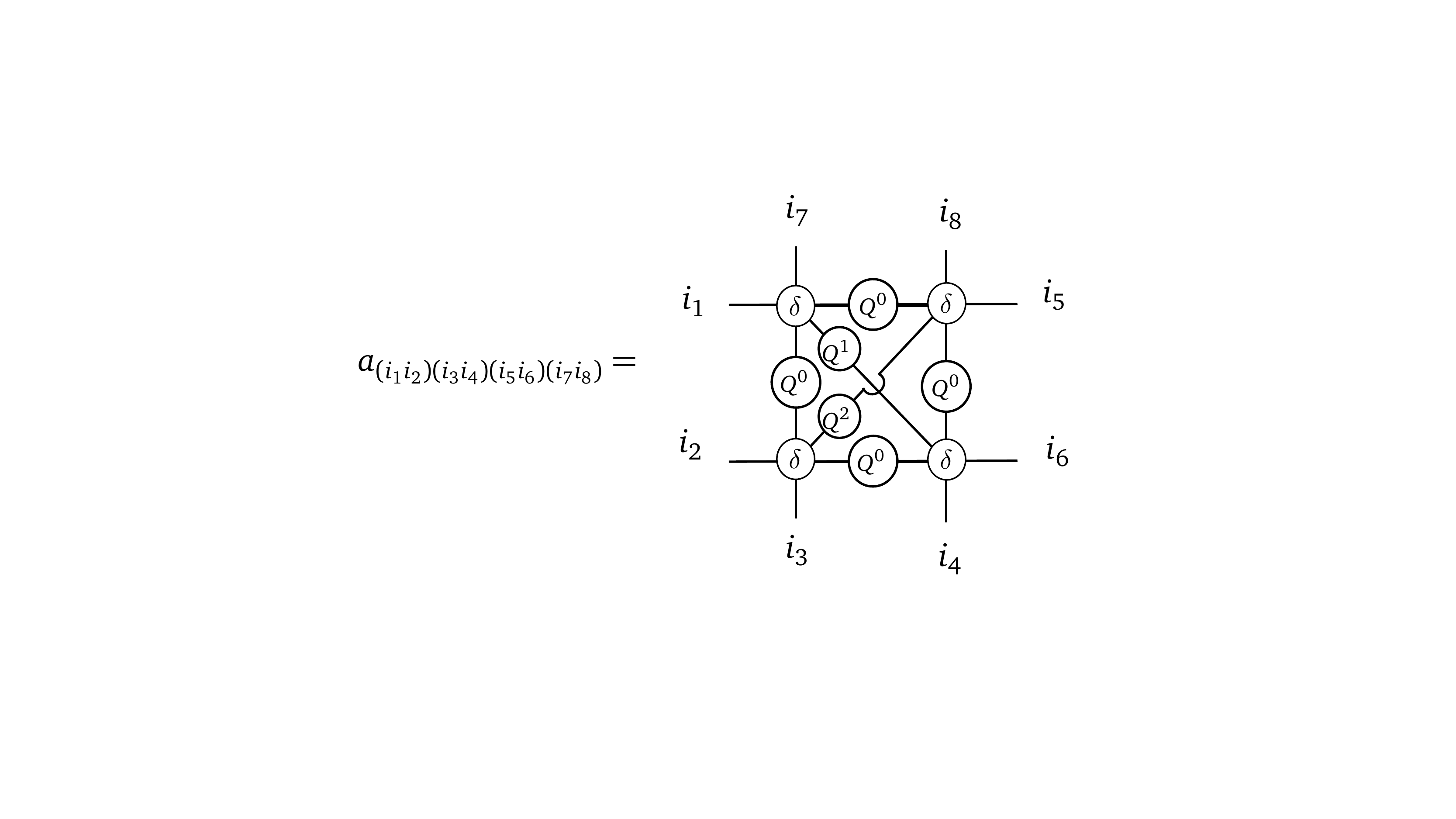} 
\caption{Definition of the local tensor $a$ representing a plaquette configuration.}
\label{fig:litta}
\end{figure}

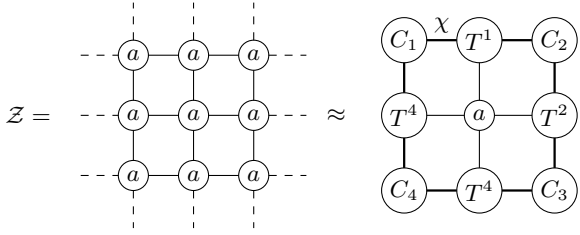
\begin{figure}
\begin{center}
\begin{tikzpicture}
\draw (-5, -1) node{$\mathcal{Z}= $};
\draw (-0.9, -1) node{$\approx$};
\draw (0.5, 0.2) node{$\chi$};
\draw (-3.6, -0.2) circle (0.2cm) node{$a$}; \draw (-2.8, -0.2) circle (0.2cm) node{$a$}; \draw (-2, -0.2) circle (0.2cm) node{$a$}; 
\draw (-3.6, -1) circle (0.2cm) node{$a$}; \draw (-2.8, -1) circle (0.2cm) node{$a$}; \draw (-2, -1) circle (0.2cm) node{$a$}; 
\draw (-3.6, -1.8) circle (0.2cm) node{$a$}; \draw (-2.8, -1.8) circle (0.2cm) node{$a$}; \draw (-2, -1.8) circle (0.2cm) node{$a$}; 
\draw  [dashed] (-4.3, -0.2)--(-3.8, -0.2);\draw (-3.4, -0.2) -- (-3, -0.2); \draw (-2.6, -0.2)--(-2.2,-0.2); \draw [dashed] (-1.8,-0.2) -- (-1.3,-0.2);
\draw  [dashed] (-4.3, -1)--(-3.8, -1);\draw (-3.4, -1) -- (-3, -1); \draw (-2.6, -1)--(-2.2,-1); \draw [dashed] (-1.8,-1) -- (-1.3,-1);
\draw  [dashed] (-4.3, -1.8)--(-3.8, -1.8);\draw (-3.4, -1.8) -- (-3, -1.8); \draw (-2.6, -1.8)--(-2.2,-1.8); \draw [dashed] (-1.8,-1.8) -- (-1.3,-1.8);
\draw [dashed] (-3.6, 0)--(-3.6, 0.5); \draw [dashed](-2.8, 0)-- (-2.8, 0.5); \draw [dashed](-2, 0)-- (-2, 0.5);
\draw (-3.6, -0.4)-- (-3.6, -0.8); \draw (-2.8, -0.4)-- (-2.8, -0.8); \draw (-2, -0.4)-- (-2, -0.8);
\draw (-3.6, -1.2)-- (-3.6, -1.6); \draw (-2.8, -1.2)-- (-2.8, -1.6); \draw (-2, -1.2)-- (-2, -1.6);
\draw [dashed] (-3.6, -2)--(-3.6, -2.5); \draw [dashed](-2.8, -2)-- (-2.8, -2.5); \draw [dashed](-2, -2)-- (-2, -2.5);
\draw (0,0) circle (0.3cm) node{$C_1$}; \draw (1,0) circle (0.3cm) node{$T^1$}; \draw (2,0) circle (0.3cm) node{$C_2$};
\draw (0,-1) circle (0.3cm) node{$T^4$}; \draw (1, -1) circle(0.2cm) node{$a$}; \draw (2,-1) circle (0.3cm) node{$T^2$};
\draw (0,-2) circle (0.3cm) node{$C_4$}; \draw (1,-2) circle (0.3cm) node{$T^4$}; \draw (2,-2) circle (0.3cm) node{$C_3$};
\draw [line width=0.3mm] (0.3,0) --  (0.7,0); \draw [line width=0.3mm] (1.3,0) -- (1.7,0);
\draw [line width=0.3mm] (0,-0.3)--(0,-0.7); \draw (1,-0.3) -- (1,-0.8); \draw  [line width=0.3mm] (2,-0.3) -- (2,-0.7);
\draw (0.3,-1) -- (0.8,-1); \draw (1.7,-1) -- (1.2,-1);
\draw [line width=0.3mm] (0,-1.3) -- (0,-1.7);  \draw (1,-1.7) -- (1, -1.2);\draw [line width=0.3mm] (2,-1.3) -- (2,-1.7);
\draw  [line width=0.3mm] (1.7,-2) -- (1.3,-2); \draw   [line width=0.3mm](0.7,-2) -- (0.3,-2);
\end{tikzpicture}
\caption{Partition function first written as an infinite tensor network and then contracted to an environment made of 8 tensors. The thin lines represent bonds of dimension 4 while bold lines have dimension $\chi$.}
\label{fig:PartitionFunction}
\end{center}
\end{figure}

Although mainly used nowadays for contraction of wave-function in two dimensional quantum systems, the corner transfer matrix algorithm was first introduced by Nishino and Okunishi\cite{nishino} as a combination of Baxter's corner transfer matrices\cite{BaxterBook} and Steve White's density matrices\cite{dmrg1} in the context of two dimensional classical system as a contraction algorithm for partition functions. Indeed, one can write the partition function in the thermodynamic limit as an infinite tensor network of local tensor $a$ as shown in Fig. \ref{fig:PartitionFunction}. Multiple choices of the local tensor $a$ exist and we choose this tensor such that it describes a plaquette rather than a site. We show the diagrammatic representation of $a$ in Fig. \ref{fig:litta}.  The Boltzmann weights $Q^i$ are defined by
\begin{align}
	Q^0 = \begin{pmatrix}
		0 & 1 \\
		1 & 1
	\end{pmatrix},
	\quad
		Q^1 = \begin{pmatrix}
		e^{L} & 1 \\
		1 & 1
	\end{pmatrix}, \text{ and } 
	\quad
	Q^2 = \begin{pmatrix}
		e^{M} & 1\\
		1 & 1
	\end{pmatrix}
\end{align}
where $Q^0$ represents the hard-core constraint while $Q^2$ and $Q^1$ represent respectively the diagonal and anti-diagonal interactions. And $\delta$ takes the value $e^{\mu/4}$ if all indices are equal to one and zero otherwise. Written in this way, the tensor $a$ is of dimension $4\times 4\times 4\times 4$.

\begin{figure}
\begin{center}
\begin{tikzpicture}
\draw (0,0) circle (0.3cm) node{$C_1$}; \draw (1,0) circle (0.3cm) node{$T^1$};
\draw (0,-1) circle (0.3cm) node{$T^4$}; \draw (1,-1) circle (0.2cm) node{$a$};
\draw [line width=0.3mm] (0.3, 0) -- (0.7,0); \draw  [line width=0.3mm](1.3,0)--(1.5,0);
\draw [line width=0.3mm] (0,-0.3) -- (0,-0.7); \draw (1,-0.3)--(1, -0.8);
\draw  (0.3, -1) -- (0.8,-1); \draw (1.2,-1)--(1.5,-1);
\draw [line width=0.3mm] (0,-1.3)-- (0,-1.5); \draw (1,-1.2)--(1,-1.5);
\draw (1.5,0.1) -- (1.5,-1.1) -- (1.7,-1.1) -- (1.7,0.1) -- (1.5,0.1);
\draw [line width=0.3mm] (1.7, -0.5)--(1.9, -0.5);
\draw (1.9, -0.3) node{$\mathcal{U}_1$};
\draw (-0.1,-1.5) -- (-0.1,-1.7) -- (1.1,-1.7) -- (1.1,-1.5) -- (-0.1,-1.5);
\draw [line width=0.3mm] (0.5, -1.7)--(0.5, -1.9);
\draw (0.7, -1.9) node{$\mathcal{U}_4^\dagger$};
\draw [line width=0.3mm] (4.5,0)--(4.7,0);  \draw (5,0) circle (0.3cm) node{$T_1$}; \draw [line width=0.3mm] (5.3,0)--(5.5,0); 
\draw (5,-0.3)--(5,-0.8);
\draw (4.5,-1)--(4.8,-1);  \draw (5,-1) circle (0.2cm) node{$a$}; \draw (5.2,-1)--(5.5,-1); 
\draw (5,-1.2)--(5,-1.5);
\draw (5.5, 0.1) -- (5.5,-1.1) -- (5.7, -1.1) -- (5.7,0.1) -- (5.5, 0.1);
\draw [line width=0.3mm] (5.7,-0.5)--(5.9,-0.5);
\draw (5.9, -0.3) node{$\mathcal{U}_1$};
\draw (4.5, 0.1) -- (4.3,0.1) -- (4.3, -1.1) -- (4.5,-1.1) -- (4.5, 0.1);
\draw [line width=0.3mm] (4.1,-0.5)--(4.3,-0.5);
\draw (4.1, -0.3) node{$\mathcal{U}_1^\dagger$};
\end{tikzpicture}
\caption{Full iteration for the corner tensor $C_1$ and row tensor $T_1$. Other tensors are grown similarly.}
\label{fig:iteration}
\end{center}
\end{figure}
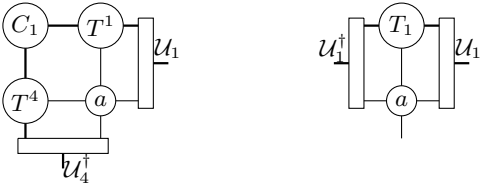

The CTMGR algorithm contracts the infinite tensor network into an environment of 8 different tensors $E = \{C_1,T_1,C_2,T_2,C_3,T_3,C_4,T_4\}$ with corner tensors $C_i$ of dimension $\chi\times \chi$ and row/ column tensors $T_i$ of dimension $\chi\times 4\times\chi$. The parameter $\chi$ is referred to as the bond dimension. In the infinite bond dimension limit, one recovers the exact result. Thus, $\chi$ controls the approximation of the algorithm. We show in Fig \ref{fig:PartitionFunction} the partition function written as a contraction of the environment and of the local tensor. When divided by the partition function, this environment becomes a measure over observables defined on the unit cell $a$. CTMRG thus gives an easy way to compute expectations of local observables. It works through a two-step iterative process referred to as {\it extension and truncation}\cite{orus} which we describe below and illustrate in Fig. \ref{fig:iteration}. \\
\textbf{Extension:} In order to increase the number of sites, to each corner tensor one adds a column, a row, and a local tensor $a$. Similarly, to each column and row tensor one adds a local tensor. One is then left with corner tensors of dimension $4\chi \times 4\chi$ and row, column tensors of dimension $4\chi \times 4\times 4\chi$. \\
\textbf{Truncation:} If the extension was repeated unchecked, the dimensions of the tensors would grow exponentially. Thus, one needs to project the tensor in a relevant subspace. Such projectors are commonly denoted as isometries and are computed through the singular value decomposition of some density matrices. Multiple choices of density matrices are possible and will lead to different convergence. We choose the one originally introduced by Nishino and Okunishi as
\begin{align}
 \mathcal{U}_1' \mathcal{S} \mathcal{V}_1 & = C_2' C_3' C_4' C_1' \\
 \mathcal{U}_2' \mathcal{S} \mathcal{V}_2 & = C_3' C_4' C_1' C_2' \nonumber \\
 \mathcal{U}_3' \mathcal{S} \mathcal{V}_3 & = C_4' C_1' C_2' C_3' \nonumber \\
 \mathcal{U}_4' \mathcal{S} \mathcal{V}_4 & = C_1' C_2' C_3' C_4' \nonumber
\end{align}
where the isometries $\mathcal{U}_i$ are obtained by keeping the $\chi$ largest singular values. Repeating those two steps will increase the size of the lattice until convergence, at which point one considers the thermodynamic limit to have been reached. The convergence is checked through the difference of energy per site between two iterations.

\section{Correlation length and wave-vector}

\begin{figure}[t!]
\centering
\includegraphics[width = 0.45\textwidth]{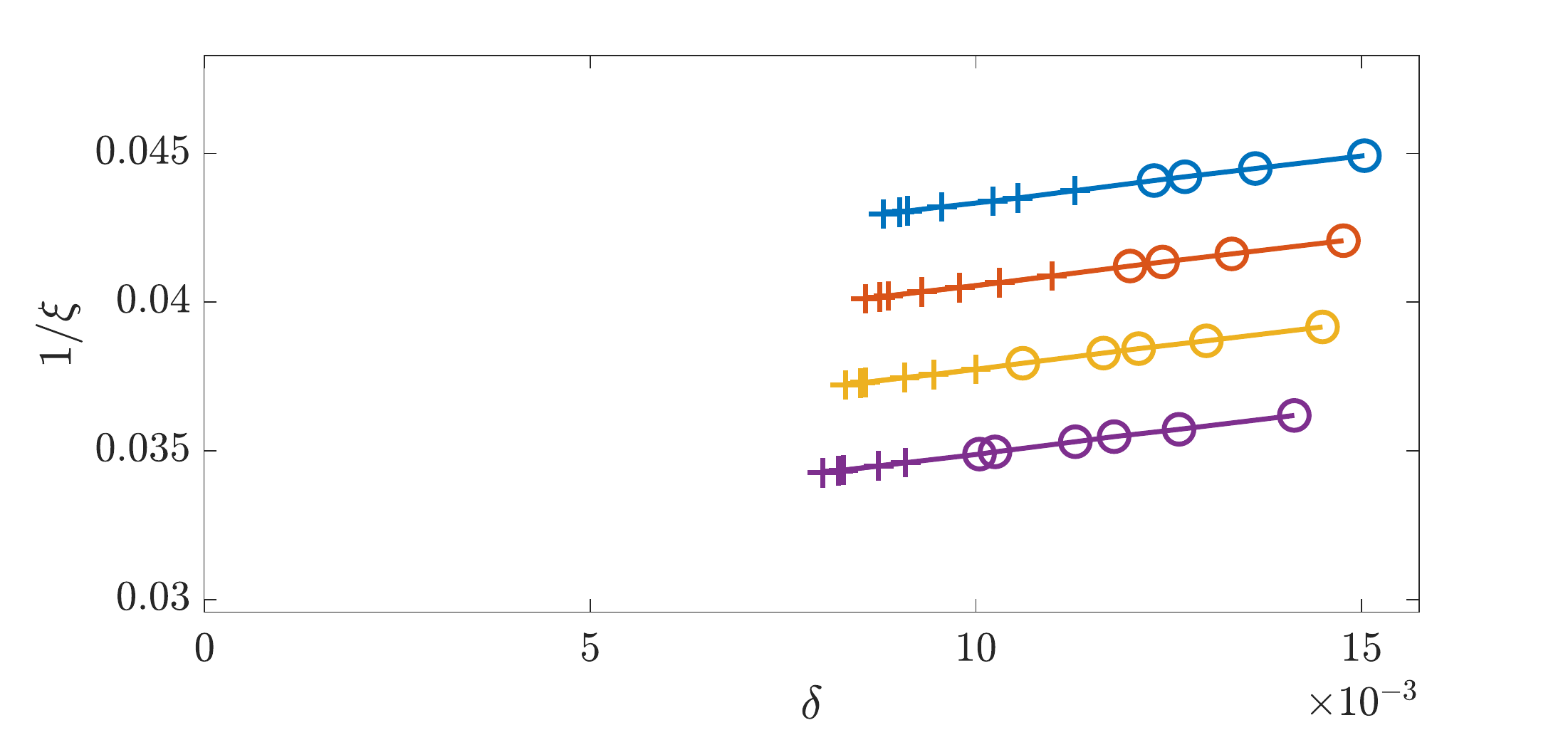}
\caption{Extrapolation of the inverse correlation length with respect to gaps in the spectrum of the transfer matrix for bond dimensions $\chi$ between 100 and 200. The simulations have been performed for $z = 1.4$ in the period-3 phase. Each colour represents a different temperature in the ordered phase. Due to a level crossing in the transfer matrix, we used two different gaps: $\delta = \epsilon_7 - \epsilon_4 (+)$  and $\delta = \epsilon_9 - \epsilon_4 (\circ)$.}
\label{fig:Extrap}
\end{figure}

The main advantage of the CTMRG method is that it gives direct access to the transfer matrices, which in turn give aceess to the correlation length and to the wave-vector. Indeed, denoting the ordered normalised eigenvalues by
\begin{align}
    \lambda_j = e^{-\epsilon_j + i\phi_j}, \qquad j=1,2,\ldots
\end{align}
one can show that the correlation length and wave-vector are given by
\begin{align}
    \frac{1}{\xi} = \epsilon_2, \qquad q = \phi_2.
\end{align}
%In order to describe two-point functions with an algebraic pre-factor, one needs the spectrum of the transfer matrix to converge to some continuum in the infinite bond dimension limit. 
%In order to accurately describe two-point functions, the spectrum of the transfer matrix above the first gap needs to converge to some continuum in the infinite bond dimension limit.
In the generic situation where the correlation decays with a power-law prefactor, the spectrum of the transfer matrix is expected to converge to a continuum above the first gap in the infinite bond dimension limit.
This was first proposed by Rham \textit{et al}\cite{czarnik2018} as a mean of extrapolation. More precisely, they suggested that the inverse correlation length behaves linearly with any gap $\delta$ of the transfer matrix as
\begin{align}
    & \epsilon_2(\chi) = 1/\xi_{\text{exact}} + b \delta. \\
\end{align}
We can define a similar extrapolation scheme for the wave-vector:
\begin{align}
    & \phi_2(\chi) = q_{\text{exact}} + b \delta' . \nonumber
\end{align}
Although in principle any gap could be used, in practice we favour smaller ones. For the incommensurate phase we systematically used
\begin{align}
\delta & = \epsilon_4 - \epsilon_2, \\
\delta' & = \phi_4 - \phi_2 . \nonumber
\end{align}
We notice that due to level crossings in the transfer matrix one might need to use different gaps for the extrapolation. We give an example of such a case in Fig. \ref{fig:Extrap}.

We note that, with  $a$ defined as in Fig. \ref{fig:litta}, the CTMRG algorithm will give the transfer matrices in the $x$ and $y$ direction. We then have access to the correlation length and wave-vectors only in those two directions. 

The error-bars on the critical exponent $\nu(T)$ are computed via a Taylor expansion as
\begin{align}
\delta \nu = \nu \left(  \frac{\delta T_c }{\mid T - T_c \mid} + \frac{\delta \xi(T)}{\xi(T) } +  \frac{\delta \xi(T-dT) + \delta \xi(T+dT)}{\xi(T+ dT) -\xi(T-dT)}  \right).
\end{align}
The error-bars on the other critical exponents are estimated in a similar way.

\section{Translational symmetry breaking} 

In the $3\times 1$ phase, the system becomes $2\pi/3$ commensurate and breaks translational invariance into three sub-lattices. As we are computing a measure over a plaquette, the algorithm will then converge to a different environment at each iteration modulo three, such that the CTMRG will converge to: $E_1 \xrightarrow{}E_2\xrightarrow[]{}E_3\xrightarrow{}E_1\xrightarrow{} \ldots $ and so on.

\begin{figure}[t!]
\centering
\includegraphics[width = 0.45\textwidth]{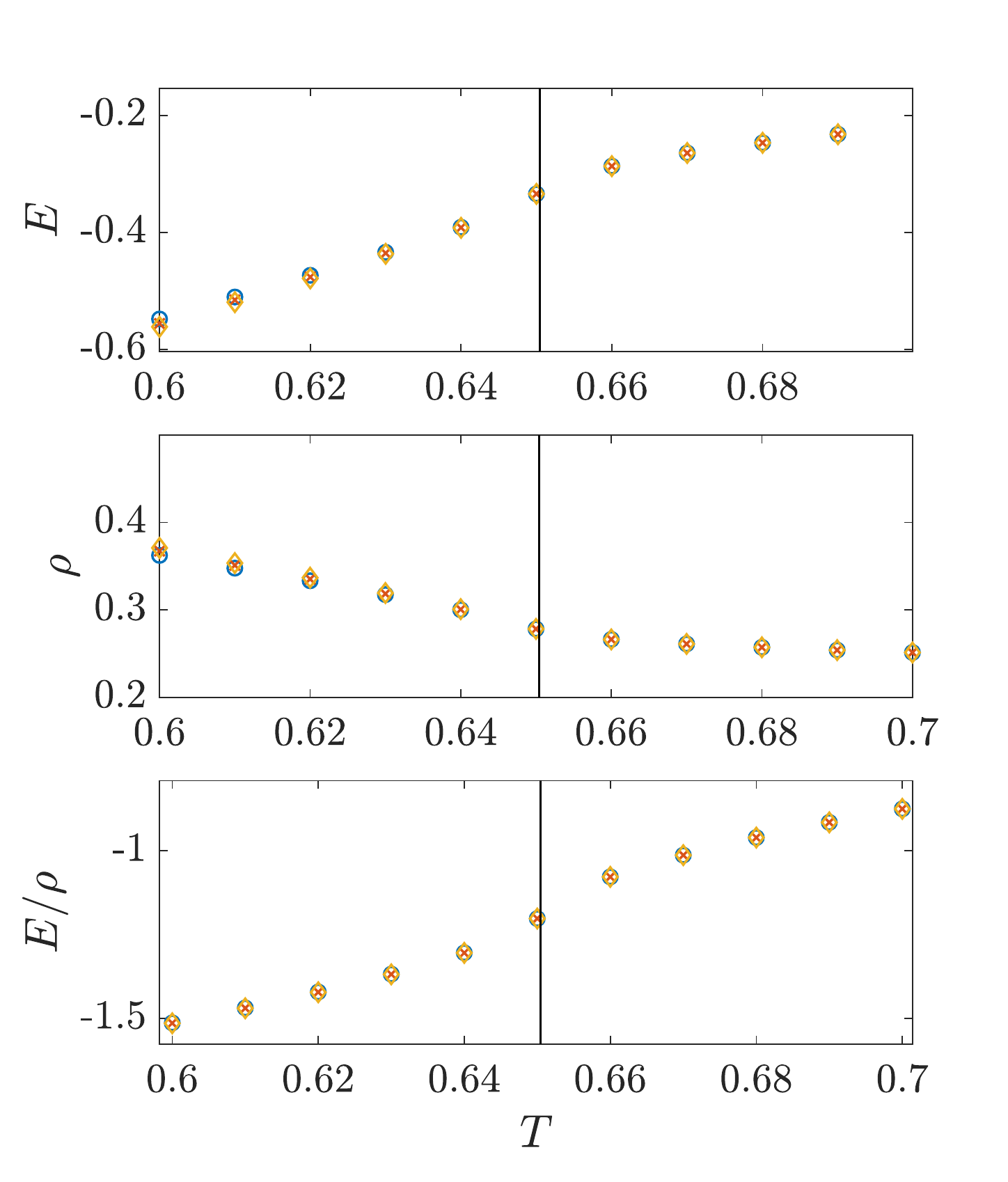}
\caption{Temperature dependence of various observables at $\chi = 50$ and at the three-state Potts critical activity $z_c$ with $M = -L$. Each colour represents the environment in which the observable was computed. We can see that in the $3\times 1$ phase, the density and the energy both depend on the environment in which they are computed. By contrast, their ratio does not. We further check that the critical density is recovered at the transition for all environments.  The vertical black lines represent the critical temperature.}
\label{fig:transl}
\end{figure}

We note that we cannot mix the different environments. Furthermore, single site operators have no reason to be equal if computed under different environments, and in general one has $\langle \mathcal{O}(x)\rangle_{E_j} \neq \langle \mathcal{O}(x)\rangle_{E_i}$ for $i\neq j$. However, the correlation length and the wave vector are expected to be independent of $E_i$. We illustrate the differences between observables in Fig. \ref{fig:transl} where the density and energy have been computed as 
\begin{align}
    &\rho  = \langle n \rangle, \\
    &  E  = \langle n_{x,y} n_{x+1, y+1}  \rangle -  \langle n_{x,y} n_{x+1, y-1}  \rangle.
\end{align}
We further notice that the ratio of the two does not depend on the environment in which it is computed as shown in Fig.\ref{fig:transl} (third panel). This can be explained by the simple fact that if a sublattice holds a larger number of particles, the absolute value of the energy increases accordingly.

 % 0.750476088529204
 % 0.750896991809627
 % 0.750747780548990

\section{Critical temperature and effective exponent}

The investigation of phase transitions and their universality classes is usually done by fitting algebraically decaying quantities close to criticality. Unfortunately, most of the time, due to various corrections and crossovers, the power law describing the latter is not exact. Hence, a naive fit will give different results depending on the parameter range used. We thus choose a different approach based on the study of effective exponents and their behaviours in the critical limit. For an algebraically diverging quantity $A\propto \mid T-T_c\mid^{\theta}$, we define its associated effective exponent as
\begin{align}
\theta_{eff} = - \frac{d\log(A)}{d\log(t)} 
\end{align}
with $t = \mid T-T_c\mid$. We note that as the temperature approaches the transition, one recovers the critical exponent. It turns out that we can distinguish between a two-step transition and a unique one simply by analysing the critical limit of the effective exponent $\nu_{eff}$. Indeed, if the transition is unique, one expects the exponents from both sides of the transition to converge to a unique value at criticality. This criterion can also be used to fix $T_c$. By contrast, if the transition is a two-step one, setting a unique limit will result in $\nu>1$ because of the Kosterlitz-Thouless nature of the transition at high temperature. Such a large value of the exponent is not expected in this model and one can conclude that the transition occurs in two steps. In that case, one expects the low temperature transition to be described by the Pokrovsky-Talapov critical exponent and we can fix $T_{PT}$ such that $\nu_{x+y}^{LT} = \bar{\beta}$ or $\nu_{x-y}^{LT} = 1$ at the transition with $LT$ denoting the exponents defined in the low temperature regime.

\section{Quantum - classical correspondence}
Consider the rescaled hard-core boson hamiltonian given by 
\begin{align*}
H = \sum_i^N - (\hat{d}_j + \hat{d}_j^\dagger) + \frac{U}{\omega} \hat{n}_j + \frac{V}{\omega} \hat{n}_j \hat{n}_{j+2}.
\end{align*}
We denote by $\mid n_j \rangle$ the eigenstate of $\hat{n}_j$ with eigenvalue $n_j$. Then the partition function is given by
\begin{align}
Z & = \text{Tr} e^{-H} \nonumber \\
& = \sum_{\{n\}} \prod_l \langle n_1^{l} \ldots n_N^{l} | e^{-\frac{H_0}{n} -\frac{H_1}{n} } +O\left(\frac{1}{n}\right) | n_1^{l+1} \ldots n_N^{l+1} \rangle 
\label{eq:Eq10}
\end{align}
with 
\begin{align}
H_0 & = \sum_i \frac{U}{\omega} \hat{n}_j + \frac{V}{\omega} \hat{n}_j \hat{n}_{j+2} \quad\text{and}\quad H_1  = -\sum_i (d_j + d_j^\dagger)  
\end{align}
where the second equality in \ref{eq:Eq10} is derived using the Trotter decomposition and becomes exact only in the $n\rightarrow \infty$ limit. We now drop the $O(1/n)$ term and consider the results to hold only in the large $n$ limit. One notes that $H_0$ has eigenstates $|n_1^l \ldots n_N^{l+1}\rangle$ and the terms in \ref{eq:Eq10} thus become

\begin{align}
 & \langle n_1^{l} \ldots n_N^{l} | e^{-\frac{H_0}{n} -\frac{H_1}{n} } | n_1^{l+1} \ldots n_N^{l+1} \rangle \nonumber  \\
 & = \prod_i e^{-\frac{U}{\omega n}n_j^l - \frac{V}{\omega n} n_j^l n_{j+2}^l } \langle n_1^{l} \ldots n_N^{l} | e^{-\frac{H_1}{n} } | n_1^{l+1} \ldots n_N^{l+1} \rangle  .
\label{eq:Eq13}
\end{align}

One further notes that $\hat{d}_j + \hat{d}_j^\dagger = \sigma_j^x$ and its exponential acts as
\begin{align*}
& e^{\sigma_j^x/n}  | n_j^{l+1}  \rangle  =  \Lambda e^\gamma  | n_j^{l+1}  \rangle + \Lambda e^{-\gamma}   | 1-n_j^{l+1} \rangle 
\end{align*} 
with
\begin{align*}
\gamma & = -\frac{1}{2}\log\tanh\left( \frac{1}{n}\right),\quad \text{and} \quad \Lambda^2  = \sinh\left(\frac{1}{n}\right) \cosh\left( \frac{1}{n}\right)
\end{align*}
Thus, the overlap with $\langle n_j^{l}  |$ can be written as
\begin{align*}
& \langle n_j^{l} | e^{ \sigma_j^x/n}  | n_j^{l+1}  \rangle = \Lambda e^{\gamma (2n_j^l-1)(2n_j^{l+1} -1)}
\end{align*} 
and the product of $e^{\sigma_j^x/n}$ simply becomes
\begin{align*}
 \langle n_1^{l} \ldots n_N^{l} | \prod_i e^{ \sigma_j^x/n}   | n_1^{l+1} \ldots n_N^{l+1} \rangle  &= \Lambda^N e^{\sum_j \gamma (2n_j^l-1)(2n_j^{l+1} -1)}.
\end{align*}
Finally, combining the above equation with Eqs.\ref{eq:Eq13} and \ref{eq:Eq10} allows one to write the partition function as
\begin{align*}
Z & = \Lambda^N \sum_{\{n\}} \prod_l e^{\sum_i -\frac{U}{\omega n}n_j^l - \frac{V}{\omega n} n_j^l n_{j+2}^l + \gamma (2n_j^l-1)(2n_j^{l+1} -1)}.
\end{align*}
One recognises the diagonal transfer matrix of the Hard-square model and can identify $V,U$ and $\gamma$ with $M,L$ and $\mu$ as
\begin{align}
& -\frac{V}{\omega n} = M, \quad 4\gamma = L, \quad \log(z) = -4\gamma - \frac{U}{\omega n}.
\label{eq:Eq12}
\end{align}
In the large $n$ limit, we approximate $\tanh(\frac{1}{n}) \simeq \frac{1}{n}$ and the second equality of the above equation gives $e^{-L/2} = \frac{1}{n}$, leading to
\begin{align*}
 \frac{V}{\omega} = -M e^{L/2} .
\end{align*}
Finally, using the third equality in \ref{eq:Eq12} and a Taylor expansion around $-\frac{U}{\omega} e^{-L/2}$ gives
\begin{align*}
 z & = e^{-4\gamma}e^{-\frac{U}{\omega}e^{-L/2}} \\
 & =  e^{-L}(1 - \frac{U}{\omega}e^{-L/2}) \\
 &\rightarrow \frac{U}{\omega} = (1-ze^{L} )e^{L/2}.
\end{align*}
We recover Eq.\ref{eq:EqSF}, mapping the hard-core boson to the hard-square model. The bosonic hard-core constraint $\hat{n}_i \hat{n}_{i+1} = 0$ naturally translates into forbidding two neighbouring sites to be both filled.

\end{document}